\documentclass[twocolumn,aps,prb,floatfix,letterpaper]{revtex4}
\usepackage{graphicx,amsmath,amsfonts,amssymb,ulem}
\usepackage{comment}
\usepackage[usenames,dvipsnames]{color}
\usepackage{newfloat}
\usepackage{yquant}
\usepackage{tikz}
\usetikzlibrary{quantikz}
\usepackage[tight]{subfigure}
\usepackage[colorlinks,linkcolor=blue,urlcolor=red,citecolor=red]{hyperref}
\usepackage{textcomp}
\usepackage{physics}
\usepackage{soul}
\usepackage{tabularx}
\usepackage{mathtools}
\usepackage{float}
\usepackage{amsmath}
\usepackage{soul}
\usepackage {ctable}
\definecolor{LightCyan}{rgb}{0.88,1,1}
\usepackage{multirow}
\usepackage{newfloat}
\DeclareFloatingEnvironment[
    fileext=loa,
    listname=List of Algorithms,
    name=ALGORITHM,
    placement=tbhp,
]{algorithm}
\hypersetup{
colorlinks=true,
linkcolor=blue,
filecolor=magenta,      
urlcolor=red,
citecolor=red,
pdftitle={Frag-PBC},
}
\DeclareFloatingEnvironment[name=Box,within=none,placement=tbp]{CGBox}

\mathchardef\mhyphen="2D

\newcommand{\qket}[1]{\ket{\widetilde{#1}}}

\begin{document}

\normalem
\title{A Synthesis of Hidden Subgroup Quantum Algorithms and \\
   Quantum Chemical Dynamics}
\author{Srinivasan S. Iyengar
\footnote{Email: iyengar@indiana.edu }}
\author{Anup Kumar}
\author{Debadrita Saha}
\affiliation{Department of Chemistry, and the Indiana University Quantum Science and Engineering Center (IU-QSEC),
Indiana University, 800 E. Kirkwood Ave, Bloomington, IN-47405}
\author{Amr Sabry
\footnote{Email: sabry@indiana.edu }}
\affiliation{Department of Computer Science, School of Informatics, Computing, and Engineering, and the Indiana University Quantum Science and Engineering Center (IU-QSEC),
Indiana University, Bloomington, IN-47405}
\date{\today}

\begin{abstract}
  We describe a general formalism for quantum dynamics and show how this formalism subsumes several quantum algorithms including the Deutsch, Deutsch-Jozsa, Bernstein-Vazirani, Simon, and Shor algorithms as well as the conventional approach to quantum dynamics based on tensor networks. The common framework exposes similarities among quantum algorithms and natural quantum phenomena: we illustrate this connection by showing how the correlated behavior of protons in water wire systems that are common in many biological and materials systems parallels the structure of Shor's algorithm.
\end{abstract}

\maketitle

%%%%%%%%%%%%%%%%%%%%%%%%%%%%%%%%%%%%%%%%%%%%%%%%%%%%%%%%%%%%%%%%%
\section{Introduction}
\label{introduction}

The promise of solving complex problems efficiently using quantum computing hardware and associated software is a rapidly evolving research frontier\cite{Preskill2018-NISQ,Aspuru-Guzik-Science-2005}. While we are in the very early stages of this upcoming quantum revolution, there are a diverse set of important scientific and technological areas that may greatly benefit from such developments. One key quantum algorithm that started this entire debate approximately 40 years ago is Shor's algorithm\cite{Shor,Nielsen-Chuang}. Here, a quantum system can, in principle, factorize large integers into prime factors using ${{\cal O}\!\left((\log N)^{2}(\log \log N)(\log \log \log N)\right)}$ fast multiplications\cite{Shor}. Since this is exponentially faster than the traditional classical approach which requires ${{\cal O}\!\left(e^{1.9(\log N)^{1/3}(\log \log N)^{2/3}}\right)}$ operations, the promise of a second quantum revolution was born.

Orthogonally, the sister fields of atomic and molecular physics and quantum chemistry have learned to wonder if atoms and molecules store and propagate quantum information. While it has been known that such ``information'' indeed evolves in time as per the laws of quantum theory, one may also ask if chemical reactions and chemical transformations are, indeed, algebraic transformations that ``compute'' new information not dissimilar from quantum algorithms.
That is, is the time-evolution of a molecular processes to be interpreted as a computational protocol that is ``programmed'' by nature, or through clever use of synthetic techniques? However, the study of molecular dynamics is complicated by the fact that molecules contain many correlated degrees of freedom. For example, with ${\cal D}$ degrees of freedom and ${\cal N}$ basis representation per degree of freedom, the complexity of information grows approximately as ${\cal N}^{\cal D}$. As a result, quantum chemical dynamics is thought to be exponentially hard. 

To alleviate this rather catastrophic situation, tensor networks\cite{tensor_network,TensorTrain} have recently become popular. Tensor Networks (TN) have roots in the tensor decomposition field of multi-linear algebra \cite{Kolda2009-sb} and are a general framework for data compression and have proven to be effective for efficient representation of many-body quantum states in strongly correlated systems\cite{tensor_network,Verstraete2008-ah,Bridgeman2017-mh,Montangero2018-ct,Ran2020-ve,Silvi2019-hl,Orus2019-wn,Orus2014-gg,Biamonte2017-xj,Biamonte2019-du,Eisert2013-yn}.  While a tensor networks treatment adaptively truncates the Hilbert space based on the intrinsic entanglement  within the problem, given the advent of novel quantum computing algorithms, tensor networks have also proved to be a natural resource for developing new quantum algorithms \cite{Qubit-fragmentation,Fried2018-th,Schutski2020-rp,Bauer2020-cb}. The approach has been shown to have applications for low-energy states of local, gapped Hamiltonians which are characterized by satisfying a so called area-law of entanglement \cite{tensor_network}.
The introduction of the density matrix renormalization group (DMRG) \cite{DMRG-White,DMRG_MPS,DMRG,DMRGCI1}, was perhaps the catalyst for the excitement in the TN methodology; proving to be very useful for the simulation of one-dimensional quantum lattices \cite{Vidal2004-oh}, electronic structure calculations \cite{TNabinitio,Chan-While-MPS-MPO,T3NS,frag-TN-Anup}, approximations to vibrational states \cite{AS+TT,Nicole-TN,frag-TN-Anup} and image processing \cite{rajwade2013image,SVD_Noise_3,SVD_Noise_2,SVD_Noise_1,SVR}.

In this paper, we cast the basic structure present in a family of quantum algorithms that includes Shor's algorithm in an abstract fashion using the language of tensor networks\cite{tensor_network}. This presentation exposes parallels to more general quantum processes including those in most chemical systems.  Hence, we will then show how such an abstraction applies directly to many natural and synthetic chemical processes thus drawing a connection between existing quantum algorithms and chemical processes. 
 At the end of this exposition, we are forced to ask if natural processes exist that may represent mathematically-constructed, number-theoretic, algorithms.

Given this overarching theme, the paper is organized as follows. In Sec.~\ref{Shor}, we review the textbook presentation of Shor's algorithm and generalize it in Sec.~\ref{QCD-Shor}, using tensor networks, to arbitrary multi-partite systems. The formalism used in that generalization is our main technical result. Sec.~\ref{QCD-Shor-Circuit} derives the correspondence to generalized Shor-like circuits applicable to general quantum chemical dynamics problems. In Sec.~\ref{sec:water}, we exploit it to show that protonated water wire systems that are present in many biological environments, such as ion-channels and enzyme active sites, can be mapped to the circuit model exhibiting the same structure as the family of quantum algorithms under study. %exploit the common framework to show parallels between Shor's algorithm and a natural process involving 
Sec.~\ref{concl} concludes.

%%%%%%%%%%%%%%%%%%%%%%%%%%%%%%%%%%%%%%%%%%%%%%%%%%%%%%%%%%%%%%%%%
\section{Quantum Algorithms} 
\label{Shor}

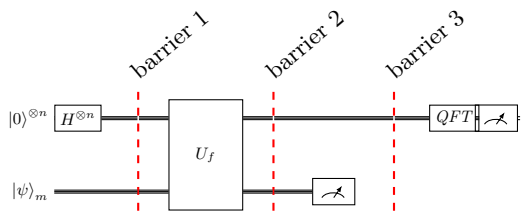
\begin{figure}[t]
  \begin{tikzpicture}[scale=0.7,every label/.style={rotate=40, anchor=south west}]
    \begin{yquant*}[operators/every barrier/.append style={red, thick},
        operator/minimum width=7mm,
        operator/separation=1mm,
        register/separation=10mm]
    qubits {$\ket0^{\otimes n}$} a;
    qubits {$\ket{\psi}_m$} b;
    box {$H^{\otimes n}$} a;
    ["barrier 1"]
    barrier (-);
    [x radius=7mm, y radius=7mm]
    box {$U_f$} (a,b);
    ["barrier 2"]
    barrier (-);
    measure b;
    discard b;
    ["barrier 3"]
    barrier (-);
    box {$\mathit{QFT}$} a;
    measure a;
    \end{yquant*}
  \end{tikzpicture}
\caption{\label{fig:templateQC}Template circuit for hidden subgroup problems.}
\end{figure}

Most quantum algorithms that are thought to be exponentially faster than their best known classical counterpart are algorithms for solving instances of \emph{hidden subgroup problems}. This family of algorithms includes the textbook quantum algorithms of Deutsch, Deutsch-Jozsa, Bernstein-Vazirani, Simon, and Shor~\cite{doi:10.1137/S0097539796300921,deutsch,deutschJozsa,365701,doi:10.1137/S0097539795293172,nielsen_chuang_2010,10.1145/237814.237866} and are all solved using the same approach illustrated in Fig.~\ref{fig:templateQC}. All the algorithms start by creating an equal superposition of all relevant possibilities, apply the $U_f$ block to the superposition, and analyze the result using the Quantum Fourier Transform (QFT). The $U_f$ block, often called the ``oracle,'' is uniformly defined as: \begin{equation}
  U_f(\ket{x}\ket{y}) = \ket{x}\ket{f(x) \oplus y},
\end{equation}
for the specific function $f$ of interest. The circuit template uses the QFT uniformly as the last step although---with the notable exception of Shor's algorithm---the low precision approximation of QFT (which is the Hadamard gate~\cite{aqft}) is often sufficient.

\begin{figure*}[t]
\input{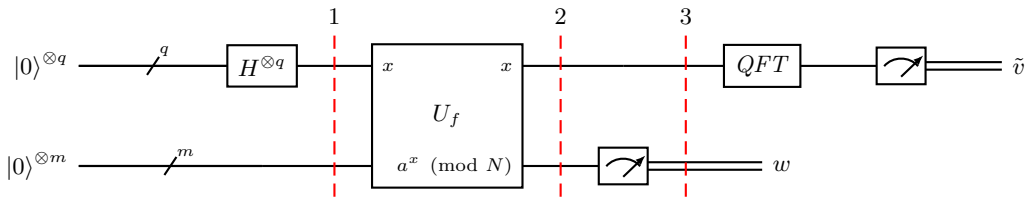}
\caption{\label{fig:shor} Quantum circuit for Shor's algorithm.}
\end{figure*}

%%%%%%%%%%%%%%%%
\subsection{Shor's Algorithm}

To be concrete, Fig.~\ref{fig:shor} instantiates the general template to the quantum circuit for an instance of Shor's algorithm for factoring the number $N$. In Stage (1), two registers are prepared: the top (input) register of $q$ qubits is initialized to an equal superposition $\frac{1}{\sqrt{2^q}} \sum_{i=0}^{2^q-1} \ket{i}$. In the bottom (output) register of $m$ qubits, each qubit is initialized to $\ket{0}$. In Stage (2), the initial state, \begin{align}
    \left\{ \frac{1}{\sqrt{2^q}} \sum_{i=0}^{2^q-1} \ket{i} \right\} \otimes \ket{0},
    \label{Shor-initial}
\end{align}
is evolved through a reversible circuit that computes $\left[ a^x \pmod{N} \right]$. The resulting state is 
\begin{align}
\frac{1}{\sqrt{2^q}}\sum_{i=0}^{2^q-1} \ket{i} \otimes \left\vert {a^i\pmod{N}} \right\rangle.
\label{Shor-final}
\end{align}
Equation (\ref{Shor-final}) represents a highly entangled state. The degree to which the two sets of registers above is entangled, is probed by computing the Schmidt number from a tensor network decomposition of the unitary evolution operations that lead to Eq. (\ref{Shor-final}), as detailed in Ref. \onlinecite{Dumitrescu-TTN-QC}. 

At Stage (3), a measurement of the output register produces some value $w$; this measurement collapses the input register to a superposition of these states $\ket{i}$ where $a^i \pmod{N} = w$. Let the number of those states be $W$; the input register state is then $\frac{1}{\sqrt{W}}\sum_{i=0}^{W-1} \ket{i}$ for those states $\ket{i}$ whose mapping by the function $a^x \pmod{N}$ produces the same value $w$. Since the function $a^x\pmod{N}$ is periodic, all these states are guaranteed to be of form $\ket{a+ks}$ for some starting offset $a$ and some multiple $k$ of the period $s$. Put differently, the state of the input register~is
\begin{align}
\sqrt{\frac{s}{W}} \sum_{k=0}^{(W/s)-1} \ket{a+ks}. 
\end{align}
It is important to note that a different measurement $w'$ of the output register would only change the starting offset~$a$ and the total number of states $W$ in the superposition, but it would \emph{not} change the period $s$. Critically, the QFT is largely insensitive to the starting offset and to the total number of states in the superposition. Its main effect is to transform a superposition of periodic states $\ket{a+ks}$ to states in the Fourier basis $\qket{v}$ such that~$v$ is close to a multiple of $W/s$. When the period $s$ is a power of 2, the Fourier states are perfectly aligned with the multiple of $W/s$ as shown by the formula below:
\begin{multline}
    \mathit{QFT}\left(\sqrt{\frac{s}{W}} \sum_{k=0}^{(W/s)-1} \ket{a+ks}\right) = \\
  \frac{1}{\sqrt{s}} \sum_{m=0}^{s-1} e^{i(2\pi/s)ma} \qket{mW/s}
\end{multline}
When the period is not a power of 2, the Fourier states with the largest probabilities are the ones close to a multiple of $W/s$. From such a measurement, some classical post-processing succeeds---with high probability---in determining the period $s$ and hence the factors of the number~$N$. 

%%%%%%%%%%%%%%%%
\subsection{Factoring Examples}
\label{Shor-15}

We illustrate the algorithm for $N=15$ and $N=21$. In the first simpler example of factoring $N=15$ we proceed as follows. In a classical preprocessing step, we choose a value for $a$ that is coprime with 15 (say 2), calculate the needed number of qubits $q=4$ and $m=4$, and generate the modular exponentiation circuit for $f(x) = 2^x \mod{15}$ using adders and multipliers~\cite{PhysRevA.54.147}. The execution of the quantum circuit proceeds as follows. The input register is initialized to the (unnormalized) equal superposition of $\ket{0} + \ket{1} + \cdots + \ket{15}$. At barrier (2), the two registers are entangled producing the (unnormalized) state $\ket{0}\ket{1} + \ket{1}\ket{2} + \ket{2}\ket{4} + \ket{3}\ket{8} + \ket{4}\ket{1} + \cdots + \ket{15}\ket{8}$. A measurement of the output register may produce $1$, $2$, $4$, or $8$ with equal probability. Say we measure $4$. The input register then collapses to the (unnormalized) state $\ket{2} + \ket{6} + \ket{10} + \ket{14}$. The QFT of this state is $\qket{0} + \qket{4} + \qket{8} + \qket{12}$. Say we measure $\qket{12}$. By properties of the QFT, we know that 12 is a multiple $m$ of $16/s$ where $s$ is the period we seek, i.e., $12m = 16/s$ or $12/16 = m/s$. The idea is that $m/s$ is guaranteed to be a small irreducible fraction that approximates $12/16$. In this case, we get the exact approximation $3/4$ from which we infer that the period is 4. From the period we calculate the two factors of 15 using $\mathrm{gcd}(15, a^{r/2} \pm 1)$ i.e., $\mathrm{gcd}(15, 3) = 3$ and $\mathrm{gcd}(15, 5) = 5$.

We follow the previous development with $N=21$, $a=10$, $q=5$, and $m=5$. At barrier (2), the (unnormalized) state is $\ket{0}\ket{1} + \ket{1}\ket{10} + \ket{2}\ket{16} + \ket{3}\ket{13} + \ket{4}\ket{4} + \ket{5}\ket{19} + \ket{6}\ket{1} + \cdots + \ket{31}\ket{10}$. Say we measure 13 at the output register. The input register collapses to the (unnormalized) state $\ket{3} + \ket{9} + \cdots + \ket{27}$. The QFT is not as perfect this time. We get the following distribution:
\begin{itemize}
\item $\qket{0}, \qket{16}$ with probability 16\%
\item $\qket{5}, \qket{11}, \qket{21}, \qket{27}$ with prob. 11\%
\item other states with negligible probabilities.
\end{itemize}
Say we measure $\qket{27}$. We know $27m$ is close to $32/s$. Equivalently we are looking for a small irreducible fraction close to $27/32$. A classical calculation produces the approximation $5/6$ yielding the period 6. From the period, we calculate $\mathrm{gcd}(21, 10^3 + 1) = 7$ and $\mathrm{gcd}(21, 10^3 - 1) = 3$.

%%%%%%%%%%%%%%%%%%%%%%%%%%%%%%%%%%%%%%%%%%%%%%%%%%%%%%%%%%%%%%%%%

%%%%%%%%%%%%%%%%%%%%%%%%%%%%%%%%%%%%%%%%%%%%%%%%%%%%%%%%%%%%%%%%%
\section{Quantum Dynamics: A Synthesis}
\label{QCD-Shor}

The quantum algorithms of Sec.~\ref{Shor} and the tensor networks approach share some apparent similarities. In both cases, the systems are composed of multiple correlated parts that evolve quantum mechanically in a system-dependent manner and that are interrogated using scenarios that measure one part of the system. This measurement affects the remaining parts of the system whose spectral properties can then be inferred using the QFT. In this section, we make this intuitive correspondence precise, opening the door for richer connections between mathematically-constructed algorithms and systems occurring in Nature.

%%%%%%%%%%%%%%%%
\subsection{Multipartite Quantum Systems} 
 
We start by considering the simple case of a system $A \oplus B$ composed of two correlated parts $A$ and $B$. Such a system can modeled in the tensor network formalism using a family of orthogonal states $\left\{ \ket{\psi^A_i} \right\}$ that may be correlated to (or entangled with) a family of mutually orthogonal bath states, $\left\{ \ket{\psi^B_i} \right\}$. The overall wavefunction is then written as a tensor product of the two components, namely, \begin{align}
    \ket{\psi} = \sum_{i,j} C_{i,j} \ket{\psi^A_i} \ket{\psi^B_j}
    \label{Tucker-2}
\end{align}
The coefficients $C_{i,j}$ capture the degree to which the parts, $A$ and $B$, are correlated to each other. For example, when $C_{i,j}$ is zero for all but one value of $i$ and~$j$, then $A$ and $B$ are completely decoupled and a product approximation suffices. However, when this is not the case, the degree to which $A$ and $B$ influence each other is often important in physical systems. Another popular example of Eq.~(\ref{Tucker-2}) are the well know Bell states\cite{Nielsen-Chuang} for two qubit systems, that are a sum of product states, \begin{align}
    \ket{\psi^1_{Bell}} = \frac{1}{\sqrt{2}} \left[ \ket{0}\ket{1} \pm \ket{1}\ket{0} \right]
    \label{Bell-1}
\end{align}
or
\begin{align}
    \ket{\psi^2_{Bell}} = \frac{1}{\sqrt{2}} \left[ \ket{0}\ket{0} \pm \ket{1}\ket{1} \right]
    \label{Bell-2}
\end{align}

Equation (\ref{Tucker-2}), may also be rewritten using Schmidt decomposition\cite{tensor_network} as, 
 \begin{align}
    \ket{\psi} = \sum_i \alpha_i \ket{\psi^A_i} \ket{\psi^B_i}
    \label{2DMPS}
\end{align}
which is essentially a bipartite Matrix Product State (MPS)~\cite{DMRG_MPS} type tensor networks\cite{DMRG-White,tensor_network} decomposition, and is usually obtained by applying a sequence of singular value decomposition steps on the Tucker-form\cite{tensor_network} of the entangled states in Eq.~(\ref{Tucker-2}). Such states are common in quantum dynamics, electronic structure, and more recently in quantum computing\cite{Qubit-fragmentation}, where the degree of correlation or entanglement between parts $A$ and $B$ are gauged using the $\left\{ \alpha_i \right\}$-values. For the so-called maximally entangled states (such as Bell-states and the Greenberger-Horne-Zellinger (GHZ) states), $\alpha_i$ is a constant value for all $i$. (Compare Eq.~(\ref{2DMPS}) with Eq.~(\ref{Bell-1}) and~(\ref{Bell-2}).) In quantum chemical dynamics such highly correlated states are not common and in general the $\alpha_i$-values may decay in some fashion when the sets of states $\left\{ \ket{\psi^A_i} \right\}$ and $\left\{ \ket{\psi^B_i} \right\}$ are appropriately ordered.
 
For a system containing ${\cal D}$ separate parts labeled as $A_{\gamma}$, $\gamma=1\cdots {\cal D}$, one may write the overall wavefunction in a form similar to Eq.~(\ref{Tucker-2}): 
 \begin{align}
    \ket{\psi} = \sum_{i,j,\cdots} C_{i,j,\cdots,{\cal D}} \ket{\psi^{A_1}_i} \ket{\psi^{A_2}_j} \cdots \ket{\psi^{A_2}_{\cal D}}
    \label{Tucker-D}
\end{align}
where $C_{i,j,\cdots,{\cal D}}$ is a rank-${\cal D}$ tensor and encodes the correlations between the constituents, $\left\{ A_{\gamma} \right\}$. The matrix product state representation of Eq (\ref{Tucker-D}), obtained from a sequence of bipartite singular value decomposition steps, yields the MPS state,
\begin{multline}
  \ket{\psi} = \\
  \sum_{\bar{i}} \ket{\psi^{A_1}_{i_1}} \beta_{i_1,i_2} \ket{\psi^{A_2}_{i_1,i_2}} \beta_{i_2,i_3} \cdots \beta_{i_{{\cal D}-1},i_{{\cal D}}} \ket{\psi^{A_{\cal D}}_{i_{{\cal D}-1},i_{{\cal D}}}} = \\
    \sum_{\bar{i}} \left[ \prod_{\gamma=1}^{{\cal D}-1} \beta_{i_\gamma,i_{\gamma+1}} \right]  
    \ket{\psi^{A_1}_{i_1}} 
    \left\{ \prod_{\gamma=1}^{{\cal D}-2} \ket{\psi^{A_\gamma}_{i_\gamma,i_{\gamma+1}}} \right\} 
    \ket{\psi^{A_{\cal D}}_{i_{{\cal D}-1},i_{{\cal D}}}}
    \label{MPS-D}
\end{multline}
where the coefficients, $\left\{ \beta_{i_\gamma,i_{\gamma+1}} \right\}$ take on a generalization of the $\alpha_i$ in Eq.~(\ref{2DMPS}) and capture entanglement in a system with ${\cal D}$ parts. 

%%%%%%%%%%%%%%%%
\subsection{Time Evolution of Eq.~(\ref{2DMPS})}

In $A \oplus B$ systems occurring in quantum dynamiccs and electron-nuclear dynamics, we are often interested in learning about the influence of each part on the other. Towards this goal, without loss of generality, we begin by introducing an initial state of the $A \oplus B$ system that is an uncorrelated bipartite simplification of Eq.~(\ref{2DMPS}), that~is,
 \begin{align}
    \ket{\psi_0} = \ket{\psi^A_0} \ket{\psi^B_0}
    \label{2DMPS-Prop-product}
\end{align}
 The time-evolution of the state $\ket{\psi_0}$ is given using a unitary evolution operator, 
\begin{align}
    {\cal U} \ket{\psi_0} = {\cal U} 
    \ket{\psi^A_0} \ket{\psi^B_0}
    \label{2DMPS-Prop}
\end{align}
which may be further explicated by writing the time-evolution operator, ${\cal U} \equiv \exp {-\imath {\cal H} t /\hbar}$ as a correlated matrix product operator\cite{tensor_network}, or a tensor product operator, 
\begin{align}
    {\cal U} = \sum_\alpha {\cal U}^A_\alpha  {\cal U}^B_\alpha 
    \label{2DMPO}
\end{align}
where, again, the multiple parts of the system are coupled by the overall Hamiltonian (and the time-evolution operator). Thus,
\begin{align}
    {\cal U} \ket{\psi_0} = 
    \sum_{\alpha} \left[ {\cal U}^A_\alpha \ket{\psi^A_0}\right] \left[{\cal U}^B_\alpha  \ket{\psi^B_0} \right] = 
    \sum_{\alpha}
    \ket{\phi^A_{\alpha}} \ket{\phi^B_{\alpha}}
    \label{2DMPS-2}
\end{align}
The structure of the Hamiltonian and the associated time-evolution operator result in the system correlations that are captured within the time-evolution process. This is represented by the sum of product states on the right side of Eq.~(\ref{2DMPS-2}). It must be noted here that while ${\cal U}$ is required to be unitary, in general, no such restrictions are present on $\left\{ {\cal U}^A_\alpha; {\cal U}^B_\alpha \right\}$. This implies that while $\ket{\phi^A_{\alpha}}$ and $\ket{\phi^B_{\alpha}}$ may not in general be normalized, the overall propagated state is always normalized. For most physical systems ${\cal U}$ has an explicit time-dependence and is given by the exponential of a Hermitian operator as noted above, and hence $\left( \ket{\phi^A_{\alpha}}; \ket{\phi^B_{\alpha}}\right) \rightarrow \left( \ket{\phi^A_{\alpha}(t)}; \ket{\phi^B_{\alpha}(t)}\right) $. In such cases the non-unitary nature of $\left\{ {\cal U}^A_\alpha; {\cal U}^B_\alpha \right\}$, combined with the unitary nature of ${\cal U}$, yields a flow of probability between parts $A$ and $B$.
 When only one term is present on the right side of Eq.~(\ref{2DMPO}), the two parts, $A$ and $B$ are uncorrelated and in such cases $\ket{\phi^A_{\alpha}(t)}$ and $\ket{\phi^B_{\alpha}(t)}$ remain  individually normalized and there is no flow of information between the two parts.
 
 Since the input states for each part of the system may in general be chosen from complete set of states, $\left\{ \ket{\psi^A_i} \right\}$ and $\left\{ \ket{\psi^B_i} \right\}$, we may expand the final states, $\ket{\phi^A_{\alpha}}$ and  $\ket{\phi^B_{\alpha}}$, using these as basis functions to obtain the  general form of the composite state after time-evolution as
 \begin{align}
    {\cal U} \ket{\psi_0} =& \sum_{\alpha} \left[ \sum_j c_{j}^{A,\alpha} \ket{\psi^A_{j}} \right] \left[ \sum_{j^\prime} c_{j^\prime}^{B,\alpha}  \ket{\psi^B_{j^\prime}} \right] \nonumber \\ =& \sum_{j,j^\prime}  \left[ \sum_{\alpha} c_{j}^{A,\alpha} c_{j^\prime}^{B,\alpha} \right] \ket{\psi^A_{j}} \ket{\psi^B_{j^\prime}}
    \label{2DMPS-gen}
\end{align}
where $c_{j}^{A,\alpha} = \bra{\psi^A_{j}} {{\cal U}}^A_\alpha \ket{\psi^A_{0}} $, and similarly for $c_{j}^{B,\alpha}$. 
Thus it is the coefficient tensor, $\sum_{\alpha} c_{j}^{A,\alpha} c_{j^\prime}^{B,\alpha}$ that builds in the correlations in Eq.~(\ref{Tucker-2}) and is obtained here through time-evolution by ${\cal U}$. 
At this stage the two parts of the system are completely correlated to the extent allowed by the propagator ${\cal U}$. In fact the extent of such a correlation may be precisely defined from the number of elements in the summation in Eqs. (\ref{2DMPO}) and (\ref{2DMPS-gen}). 

%%%%%%%%%%%%%%%%
\subsection{Final State Analysis}
\label{final-state}

At this stage in quantum chemical dynamics, there are several analysis techniques available to gauge correlations within Eq.~(\ref{2DMPS-gen}). There are two basic types of questions asked off the propagated state in Eq.~(\ref{2DMPS-gen}). In one case, it is of interest to directly Fourier transform Eq.~(\ref{2DMPS-gen}) in the time-domain since these now provide the spectroscopic signatures of the Hamiltonian, and hence the eigenspectrum of the Hamiltonian that governs the dynamics. This is common when only a few degrees of freedom are involved and tends to become prohibitive when the number of dimensions grow. Secondly, along the lines of the topic here, one is often interested in how the system, $A$, evolves and is coupled to the properties of what we will from hereon refer to as bath, $B$. This is a general problem and includes both condensed phase quantum dynamics as well as electron-nuclear dynamics. It may also include chemical and biological sensing phenomena as signified from molecular binding processes.

Correspondingly, we project Eq.~(\ref{2DMPS-gen}) onto a specific bath state $\ket{\psi^B_k}$, which is akin to performing a measurement on the bath state, to yield,

\begin{align}
    \bra{\psi^B_k} {{\cal U}} \ket{\psi_0} =& \sum_{j}  \left[ \sum_{\alpha}  c_{j}^{A,\alpha} c_{k}^{B,\alpha} \right] \ket{\psi^A_{j}} \nonumber \\ =& \sum_{\alpha} c_{k}^{B,\alpha} \ket{\phi^A_{\alpha}} 
    \label{2DMPS-gen-Proj}
\end{align}
which represents the state after measurement on the bath state, with measurement outcome,
\begin{align}
    & Tr \left[ \ket{\psi^B_k} \bra{\psi^B_k} \left\{ {\cal U} \ket{\psi_0} \bra{\psi_0} {\cal U}^\dag \right\} \right] = Tr \left[ {\left\vert \bra{\psi^B_k} {{\cal U}} \ket{\psi_0} \right\vert}^2\right] 
    \nonumber \\ &= Tr \left[ \sum_{j,j^\prime}  \left[ \sum_{\alpha,\alpha^\prime} c_{j}^{A,\alpha} {c_{j^\prime}^{A,\alpha^\prime}}^* c_{k}^{B,\alpha} {c_{k}^{B,\alpha^\prime}}^* \right] \left\{ \ket{\psi^A_{j}}\bra{\psi^A_{j^\prime}} \right\}   \right]
    \nonumber \\ &= Tr \left[ \sum_{\alpha,\alpha^\prime} c_{k}^{B,\alpha} c_{k}^{B,\alpha^\prime} \ket{\phi^A_{\alpha}} \bra{\phi^A_{\alpha^\prime}} \right]
     \nonumber \\ &= \sum_{\alpha} {\left\vert c_{k}^{B,\alpha} \right\vert}^2
    \label{2DMPS-gen-Proj-prob}
\end{align}
Thus while the result of measurement is the net probability of Eq.~(\ref{2DMPS-gen}) along $\ket{\psi^B_k}$, that is Eq.~(\ref{2DMPS-gen-Proj-prob}), the remaining state is as in Eq.~(\ref{2DMPS-gen-Proj}). 
A large number of such measurements on the bath state will yield various outcomes 
\begin{align}
    \left\{ Tr \left[ {\left\vert \bra{\psi^B_j} {{\cal U}} \ket{\psi_0} \right\vert}^2 \right]; \ket{\psi^B_j} \right\},
    \label{MeasureB}
\end{align}
where each outcome, ${\left\vert \bra{\psi^B_j} {{\cal U}} \ket{\psi_0} \right\vert}^2$, for bath state, $\ket{\psi^B_j}$, is accompanied by the system remaining in the state given in Eq.~(\ref{2DMPS-gen-Proj}). In this manner multiple measurements of $B$, yield multiple states for $A$. Fourier transform of each of these yield,
\begin{align}
    {\text{FT}} \left\{ \sum_{j}  \left[ \sum_{\alpha}  c_{j}^{A,\alpha} c_{k}^{B,\alpha} \right] \ket{\psi^A_{j}} \right\} = {\text{FT}} \left\{ \sum_{\alpha} c_{k}^{B,\alpha} \ket{\phi^A_{\alpha}} \right\}
    \label{MeasureB-stateA}
\end{align}
essentially the state of $A$, and the degree of coupling, or entanglement, between the system ($A$) and bath ($B$) state as originally captured by Eq.~(\ref{2DMPS}). 
Equations (\ref{MeasureB}) and (\ref{MeasureB-stateA}) are critical to multiple areas of metrology in physical and biological sciences. In each case the interpretation of systems $A$ and $B$ may be different. In sensing, an analyte might bind to system $B$, which collapses the system and its Fourier transform (or a linear transform) may provide information about the analyte binding to $B$. Similar aspects exist in condensed phase quantum dynamics and chemical catalysis as well. 

%%%%%%%%%%%%%%%%
\subsection{Generalized Phase Kickback}
\label{phase-kickback}

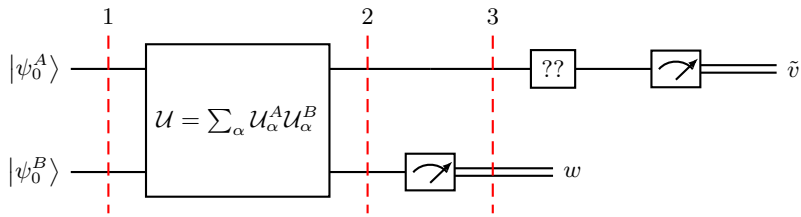
\begin{figure*}[t]
\begin{center}
\begin{quantikz}[row sep=0.7cm,column sep=1cm]
   \lstick{$\ket{\psi^A_0}$}\slice{1} & 
   \gate[wires=2][2cm]{ {\cal U} = \sum_\alpha {\cal U}^A_\alpha  {\cal U}^B_\alpha }\slice{2} &
   \qw\slice{3} &
   \gate{\mathit{??}} &
   \meter{} & 
   \rstick{$\tilde{v}$}\cw
   \\
   \lstick{$\ket{\psi^B_0}$} &
   &
   \meter{} &
   \rstick{$w$}\cw 
\end{quantikz}
\end{center}
\caption{\label{fig:shor-quantum}Quantum circuit version of bipartite quantum chemical dynamics problems. Beyond stage 3, the measurement box ``??" signifies the fact that based on different measurements, connections between known quantum algorithms and chemical dynamics problems can be established.}
\end{figure*}

In Eq.~(\ref{2DMPS-gen-Proj-prob}), part $B$ is measured by using the same basis $\left\{  \ket{\psi^B_k} \right\}$, as that used for the original propagation. Suppose this was not the case and the measurement was done using a specific ket, $ \ket{\chi^B_k}$ from within a different basis $\left\{  \ket{\chi^B_i} \right\}$ where
\begin{align}
    \ket{\chi^B_i} = \sum_j d_{i,j}^{B} \ket{\psi^B_j}
    \label{PB-1}
\end{align}
and $d_{i,j}^{B} =  \bra{\psi^B_{j}} \ket{\chi^B_i}$. In that case, Eq.~(\ref{2DMPS-gen-Proj}), takes on a more general form:
\begin{align}
    \bra{\chi^B_k} {{\cal U}} \ket{\psi_0} =& \sum_{j,j^\prime}  \left[ \sum_{\alpha} c_{j}^{A,\alpha} c_{j^\prime}^{B,\alpha} \right] \ket{\psi^A_{j}}  \bra{\chi^B_k} \ket{\psi^B_{j^\prime}} \nonumber \\ =& \sum_{j,j^\prime} d_{k,j^\prime}^{B} \left[ \sum_{\alpha} c_{j}^{A,\alpha} c_{j^\prime}^{B,\alpha}  \right] \ket{\psi^A_{j}} \label{2DMPS-gen-Proj-PK-psi} \\ =& \sum_{j^\prime,\alpha} d_{k,j^\prime}^{B}  c_{j^\prime}^{B,\alpha}   \ket{\phi^A_{\alpha}} 
    \label{2DMPS-gen-Proj-PK}
\end{align}
Equation (\ref{2DMPS-gen-Proj-PK}), as we show below, represents a generalized form of the phase kickback problem which is commonly seen in quantum information\cite{Nielsen-Chuang}. This can be illustrated by considering Eqs. (\ref{Bell-1}) and (\ref{Bell-2}) as our propagated states, ${{\cal U}} \ket{\psi_0}$. That is, to make a connection between the abstract tensor network formalism and qubits:
\begin{align}
    \left\{ \ket{\psi_i^A }\right\} \rightarrow \left\{ \ket{0}^A; \ket{1}^A \right\}
\end{align}
and similarly for $B$. Furthermore, 
\begin{align}
    \left\{ {{\cal U}} \ket{\psi_0}\right\} \rightarrow \left\{ \ket{\psi^1_{Bell}}; \ket{\psi^2_{Bell}} \right\}
\end{align}
The measurement basis for phase kickback is chosen as 
\begin{align}
    \left\{ \ket{\chi_i^B }\right\} \rightarrow \left\{ \ket{+}^B; \ket{-}^B \right\}
\end{align}
and therefore $d_{i,j}^{B} = \pm 1/\sqrt{2}$ for all $i,j$. (See Eq.~(\ref{PB-1}).)
In that case, as per Eq.~(\ref{2DMPS-gen-Proj-PK}), when measurement is constructed using $\ket{\pm^B}$, one finds,
\begin{align}
    \bra{+^B}\ket{\psi^1_{Bell}} = \frac{1}{\sqrt{2}} \ket{1}^A \pm \frac{1}{\sqrt{2}}\ket{0}^A
    \label{Bell-1-proj}
\end{align}
and 
\begin{align}
    \bra{-^B}\ket{\psi^1_{Bell}} = \frac{1}{\sqrt{2}} \ket{1}^A \mp \frac{1}{\sqrt{2}}\ket{0}^A
    \label{Bell-1-proj-2}
\end{align}
that is, system $A$ is rotated onto the $X$ basis as a result of this measurement, or the phase angle in $\bra{+^B}$ is ``kicked'' into state $A$, upon measurement. Likewise, 
\begin{align}
    \bra{+^B} \ket{\psi^2_{Bell}} = \frac{1}{\sqrt{2}}\ket{0}^A \pm \frac{1}{\sqrt{2}}\ket{1}^A
    \label{Bell-2-proj}
\end{align}
and similarly $\bra{-^B} \ket{\psi^2_{Bell}}$.

These features are captured in a general way within Eqs. (\ref{2DMPS-gen-Proj-PK-psi}) and (\ref{2DMPS-gen-Proj-PK}) where the appropriate generalization of the phase kickback is in the terms $d_{k,j^\prime}^{B} \equiv \bra{\chi^B_k} \ket{\psi^B_{j^\prime}}$, that represent the components of the measurement basis of $B$ with respect to the initial basis and also the additional basis components that are ``kicked-back'' into system $A$, as per Eq.~(\ref{2DMPS-gen-Proj-PK-psi}), after measurement on $B$.

The broader implications of Eqs. (\ref{2DMPS-gen-Proj-PK-psi}) and (\ref{2DMPS-gen-Proj-PK}) are as follows: if we consider a general system containing two entangled parts, with degree of entanglement dictated by a unitary evolution operator and hence an underlying Hamiltonian, a measurement or projection on one part, chosen as $A$ here, is also noted in $B$. Thus in some sense $B$ can ``sense'' the projection in $A$, but the extent of such a sensing process is dictated by the extent of entanglement present within ${{\cal U}} \ket{\psi_0}$ in Eq.~(\ref{2DMPS-gen}). For the Bell state the information is directly transferred whereas for the state in Eq.~(\ref{2DMPS-gen-Proj-PK-psi}), the measurement information is convoluted with the extent of entanglement.

%%%%%%%%%%%%%%%%%%%%%%%%%%%%%%%%%%%%%%%%%%%%%%%%%%%%%%%%%%%%%%%%%
\section{Circuit Model for the tensor network formalism in Section \ref{QCD-Shor}}
\label{QCD-Shor-Circuit}

The development in the previous can be recast in the circuit model to make the parallels with Shor's algorithm more explicit. We begin with Fig.~\ref{fig:shor-quantum} which provides an instance of the general formalism as a quantum circuit closely relating to the description of Shor's algorithm in Sec.~\ref{Shor}. Specifically, the initial state in Eq.~(\ref{2DMPS-Prop-product}) is chosen to be a direct product state and represents Stage~(1) in Fig.~\ref{fig:shor-quantum}. (Compare Eqs. (\ref{Shor-initial}) and (\ref{2DMPS-Prop-product}).) It must be noted that this initial state, depicted at the end of Stage~(1) in Fig.~\ref{fig:shor} was obtained from a set of Hadamard transforms that essentially provide equal weights to all components of the computational basis embedded within the first wire stream of Fig.~\ref{fig:shor}.

Thus, equivalently, $A$ in Fig.~\ref{fig:shor-quantum} may be entangled at Stage~(1), but importantly, $A$ and $B$ are uncorrelated at this initial stage and in this sense Eq.~(\ref{2DMPS-Prop-product}) resembles the initial state for all the quantum algorithms shown above.  This state is then time-evolved as dictated by Eq.~(\ref{2DMPO}) leading to Eqs. (\ref{2DMPS-2}) and (\ref{2DMPS-gen}), and represented as Stage~(2) in Fig.~\ref{fig:shor-quantum}. Similarly, the $U_f$ operator in Fig.~\ref{fig:shor} plays the same role as the propagator in Fig.~\ref{fig:shor-quantum} and presents a correlated (or entangled) state, given by Eq.~(\ref{Shor-final}) and represented at Stage~(2) in Fig.~\ref{fig:shor}. Following this time-evolution leading to Eqs. (\ref{2DMPS-2}) and (\ref{2DMPS-gen}) a measurement is constructed in all scenarios. We have only presented the analogue to Shor in Fig.~\ref{fig:shor-quantum}, complemented by the discussion in Sec.~\ref{final-state}.  The resultant state in Fig.~\ref{fig:shor-quantum}, given by Eq.~(\ref{2DMPS-gen-Proj}), now represents a general and abstract interpretation of the resultant state of Shor's algorithm, at Stage~(3), prior to the QFT step.  

The next step, as per Figs.~\ref{fig:shor} and~\ref{fig:shor-quantum}, is a Fourier transform of the state of system, $A$, as described by Eqs. (\ref{MeasureB}) and (\ref{MeasureB-stateA}), which yields the momentum representation of the resultant state in subsystem $A$. In a sense this also presents a more abstract form of the output from the top wire in Fig.~\ref{fig:shor}, and {\em one may be induced to ask if we indeed obtain a similar ``momentum representation'' for the states captured in the top wire (natural numbers) in Fig.~\ref{fig:shor}}.  Thus, at the end of this process, if the Hamiltonian represented within the propagator in Eq.~\ref{2DMPO} (or in Shor's algorithm) entangles the $A$ and $B$ dimensions, then a measurement of $B$, projects it onto a specific state.  Following this, a Fourier transform of $A$ yields the momentum representation, and in fact the power spectrum of $A$ {\em for the specific projection of $B$}.

Such a Fourier transform captures the entanglement within the composite $A \oplus B$ super-system, by probing the Fourier space structure of one part of the super-system, namely system $A$, for all possible measurement outcomes of system $B$ (assuming that multiple measurements are done on $B$).  For the specific choice of unitary in the Shor's algorithm, this Fourier spectrum of $A$ is always the same for any measurement of $B$. This may not, of course, be the case for naturally occurring or physico-chemical systems.

Finally, we also note that Equation (\ref{Tucker-2}) models the total electron-nuclear wavefunction for any molecular system that may be written as an expansion in the complete set of electronic wavefunctions with the coefficients being functions of nuclear coordinates.  In multi-dimensional correlation spectroscopy vibrational mode coupling may be studied using similar partitioning schemes where bath variables {\em influence} the dynamics of a chosen system. In fact, the example chosen in the next section is related to problems in multi-dimensional correlation spectroscopy. In chemical catalysis, ligands that surround an active site may influence the reactive process in a very strong way. Such ideas are commonly used in catalyst design. In chemical sensing of atmospheric and biological analytes, a perturbation to part $A$ through a chemical binding process, may result in a change in the state of $B$ given the extent of correlation in Eq.~(\ref{Tucker-2}).  In all such cases, one is always interested in the role that, for example, subsystem $B$ in Eq.~(\ref{Tucker-2}), plays in influencing the state of the remaining parts that are enclosed within $A$.  The resulting analysis allows one to probe the correlations between subsystems $A$ and $B$ and has numerous practical applications in the above listed set of examples.

\section{Applications: Protonated Water Wire Systems}
\label{sec:water}

We now exploit the formalism presented above to explore parallels between quantum algorithms and natural phenomena in physical, chemical, and biological systems.

%%%%%%%%%%%%%%%%
 \begin{figure}
\subfigure[]{\includegraphics[width=0.6\columnwidth]{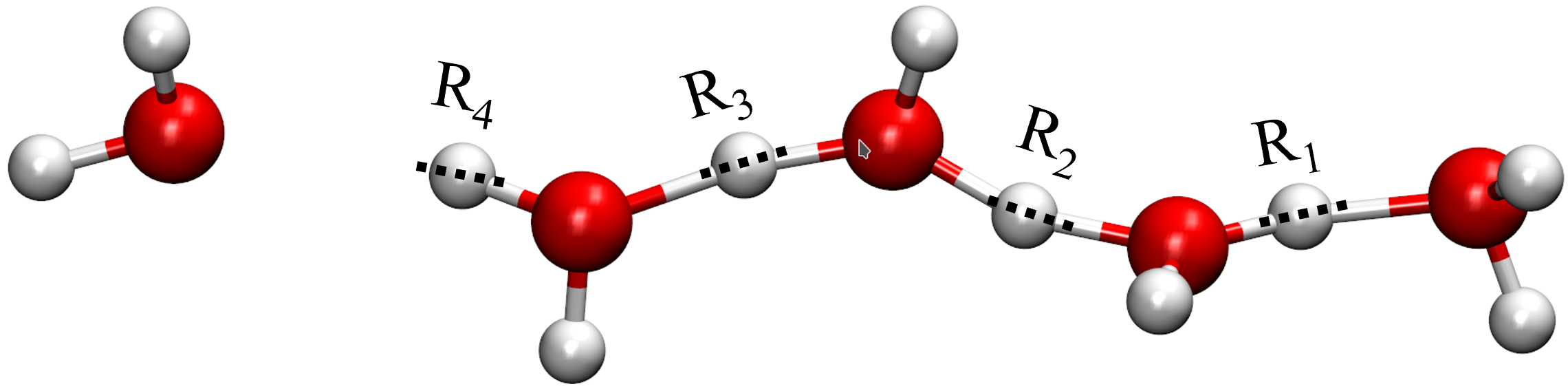}} \hspace{0.5cm}
     \subfigure[]{\includegraphics[width=0.29\columnwidth]{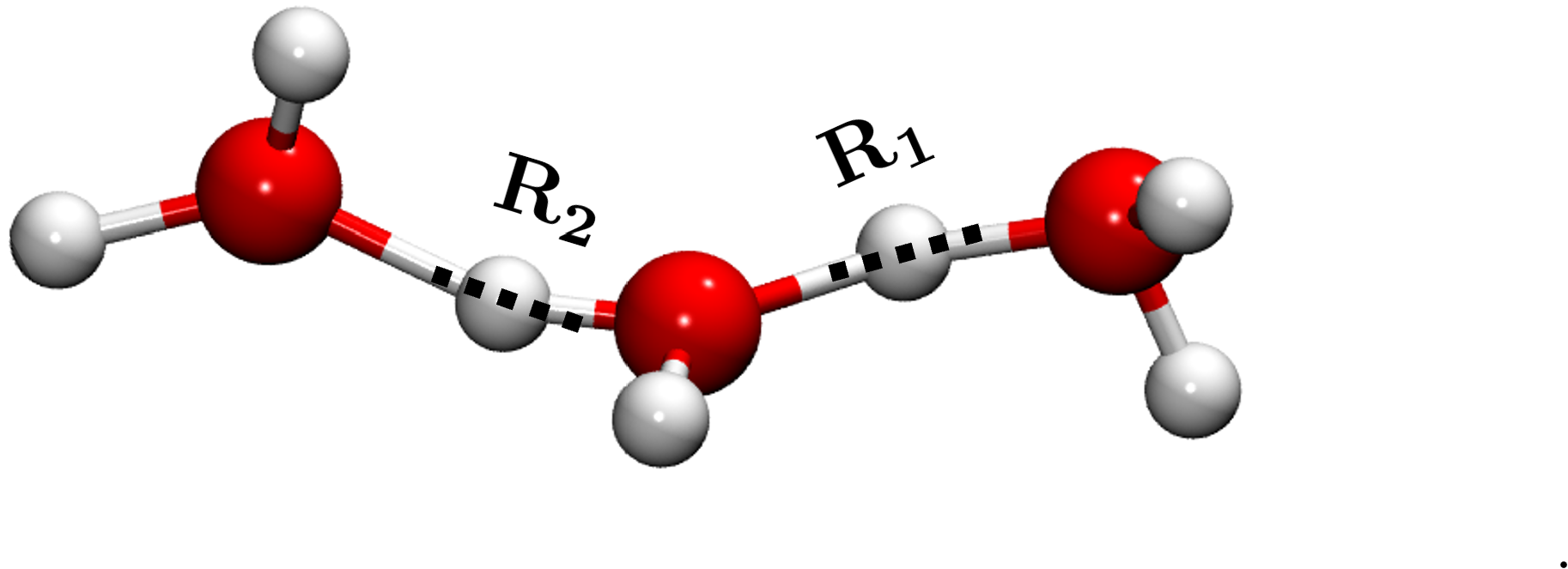}}
     \caption{\label{fig:waterwire}A protonated water wire with shared protons that are treated quantum mechanically along the grid dimensions shown. }
 \end{figure}

Protonated water wires such as those in Figs.~\ref{fig:waterwire}(a) and~\ref{fig:waterwire}(b) are encountered in a large variety of biological ion-channels, catalytic sites, light-harvesting systems, fuel cells, and form the central part of many condensed-phase chemical processes. Such systems are found in confined media, such as ion-channels. Quantum effects play a critical role in contributing to the rate of proton transport\cite{Schumaker:2000,Schumaker:2001} and in determining vibrational properties\cite{Johnson-Jordan-21mer,admp-21mer}.

\subsection{Coupled proton stretch modes in protonated water dimer}
\label{dimer}
In this section, we provide a tensor network description of such protonated systems beginning with the simple case of two protons and extending to a longer water wire chain of multiple shared protons.

\begin{figure*}
\subfigure[]{\includegraphics[width=0.32\textwidth]{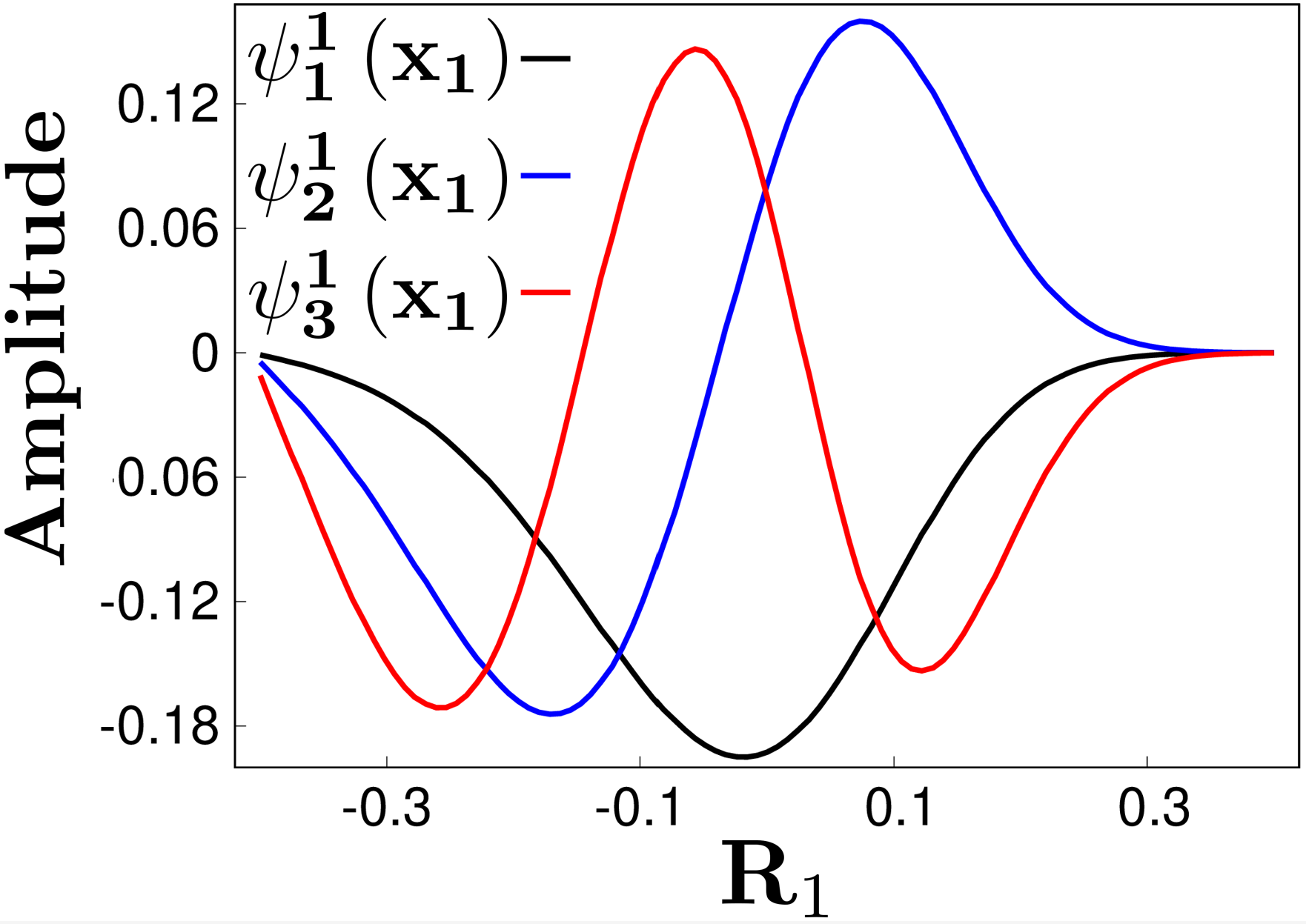}}
\subfigure[]{\includegraphics[width=0.32\textwidth]{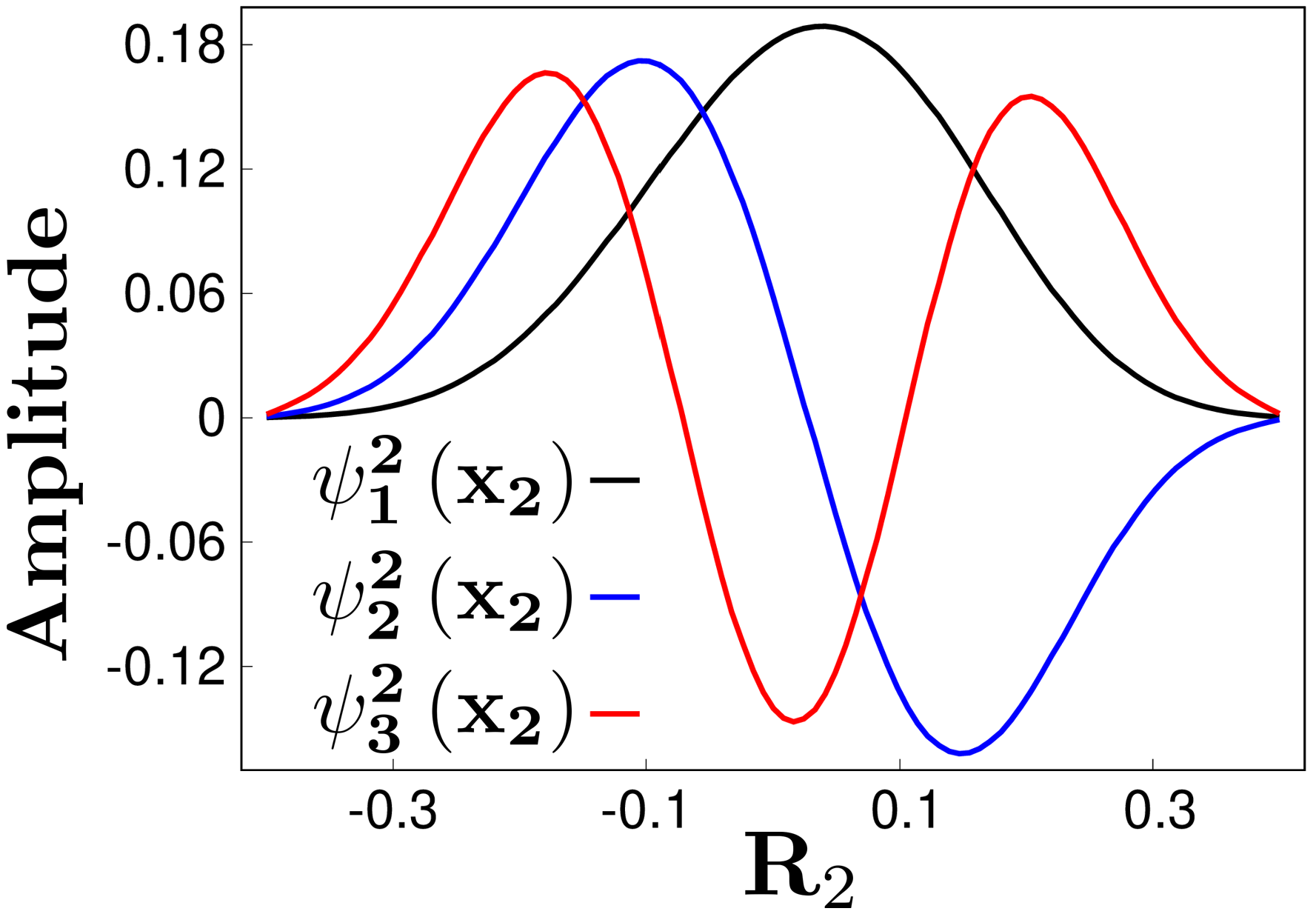}}
\subfigure[]{\includegraphics[width=0.32\textwidth]{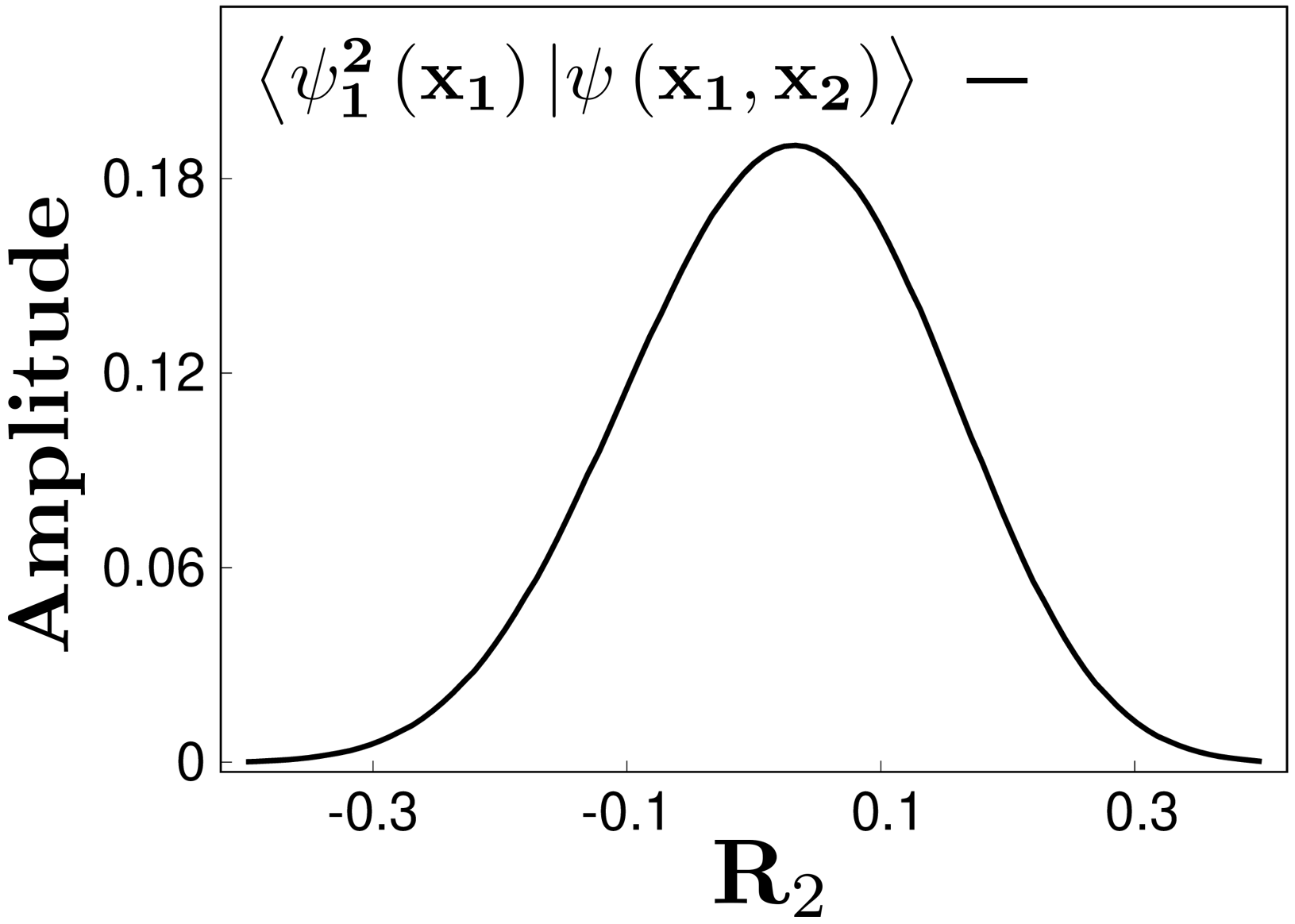}}
 \caption{Figs. (a) and (b) show functions $\left\{ \psi_i^1(x_1) \right\}$ and  $\left\{ \psi_2^1(x_2) \right\}$ that form the state in Eq.~(\ref{2D_TN_eqn}) where $\alpha_1$=0.9987, $\alpha_2$=0.0502, and $\alpha_3$=0.0022. When measured with $\psi_1^{2}(x_2)$, one obtains Fig. (c), with probability given by Eq.~(\ref{2D_TN_eqn-proj}).}
 \label{fig:2D_ww_shor}
\end{figure*}

 We begin with an analysis of the two shared proton dimensions marked as $R_1$ and $R_2$ in Fig.~\ref{fig:waterwire}(b). Using Schmidt decomposition\cite{tensor_network} the wavefunction for this two-dimensional system may be written as
\begin{equation}
     \psi(x_1,x_2) = \sum_i \alpha_i \psi_i^1(x_1)  \psi_i^{2}(x_2) \label{2D_TN_eqn}
\end{equation}
Here $\left\{ \psi_i^1(x_1) \right\}$ represents a family of functions that depicts the distribution of dimension $R_1$; this family is coupled to the family of functions, $\left\{ \psi_i^2(x_2) \right\}$ that depicts the distribution of dimension $R_2$. Additionally, $\bra{\psi_i^1}\ket{\psi_j^1} = \bra{\psi_i^2}\ket{\psi_j^2} = \delta_{i,j}$ and thus these functions form an independent orthonormal basis for the two separate dimensions. For a ${\left(\text{H}_\text{2}\text{O}\right)}_3\text{H}^+$ water wire sub-system with degrees of freedom $R_1$ and $R_2$, the wavefunction components are calculated as shown in Figs.~\ref{fig:2D_ww_shor}(a) and~\ref{fig:2D_ww_shor}(b). 

Several techniques can be used to interrogate such systems. A basic one is to perform a measurement on one of the dimensions, say $R_1$. This interrogation could be done by projecting the system onto a specific state of $R_1$, say, $\psi_k^{1}(x_2)$. We then obtain,
\begin{equation}
  \bra{\psi_k^{1}(x_2)}\ket{\psi(x_1,x_2)} = \alpha_k \psi_k^2(x_1)  \label{2D_TN_eqn-proj}
\end{equation}
based on the orthogonality conditions stated above. The associated measurement outcome is,
\begin{equation}
     {\left\vert \alpha_k \right\vert}^2 {\left\| \psi_k^2 \right\|}^2 
     \label{2D_TN_eqn-proj-prob}
\end{equation}
and for $k=1$, this quantity is close to 1 as indicated by the values of $\left\{ \alpha_i \right\}$ provided in the caption for Fig.~\ref{fig:2D_ww_shor}. 

More generally, we could also envision a more sophisticated measurement where the interrogation could be performed using the state   $\chi^{1}(x_1) \equiv 1/\sqrt{2} \left[ \psi_1^{1}(x_1) + \psi_2^{1}(x_1) \right]$ leading to a final state:
\begin{equation}
     \bra{\chi^{1}(x_1)}\ket{\psi(x_1,x_2)} = \frac{1}{\sqrt{2}} \left[ \alpha_1 \psi_1^2(x_2) + \alpha_2 \psi_2^2(x_2) \right]
     \label{2D_TN_eqn-proj-pkb}
\end{equation}
In this case, the superposition in the interrogation state is transferred to the outcome
\begin{equation}
     \frac{1}{2}\left[ {\left\vert \alpha_1 \right\vert}^2 {\left\| \psi_1^2 \right\|}^2 + {\left\vert \alpha_2 \right\vert}^2 {\left\| \psi_2^2 \right\|}^2 \right]
     \approx 0.5. 
     \label{2D_TN_eqn-proj-pkb-prob}
\end{equation}

In both interrogation scenarios, the measurement of one dimension influences the other. Thus, the form of the state in Eq.~(\ref{2D_TN_eqn}) has a significant impact on the result of the observation. If one of the states that are already within the family included in Eq.~(\ref{2D_TN_eqn}) is used as the measurement basis, the outcome is the corresponding state within that family. However, when a combination of states is used as the measurement basis, the phase kickback mechanism causes the complex phase included in this combination to make its appearance in the measured outcome.

 \begin{figure}    \includegraphics[width=\columnwidth]{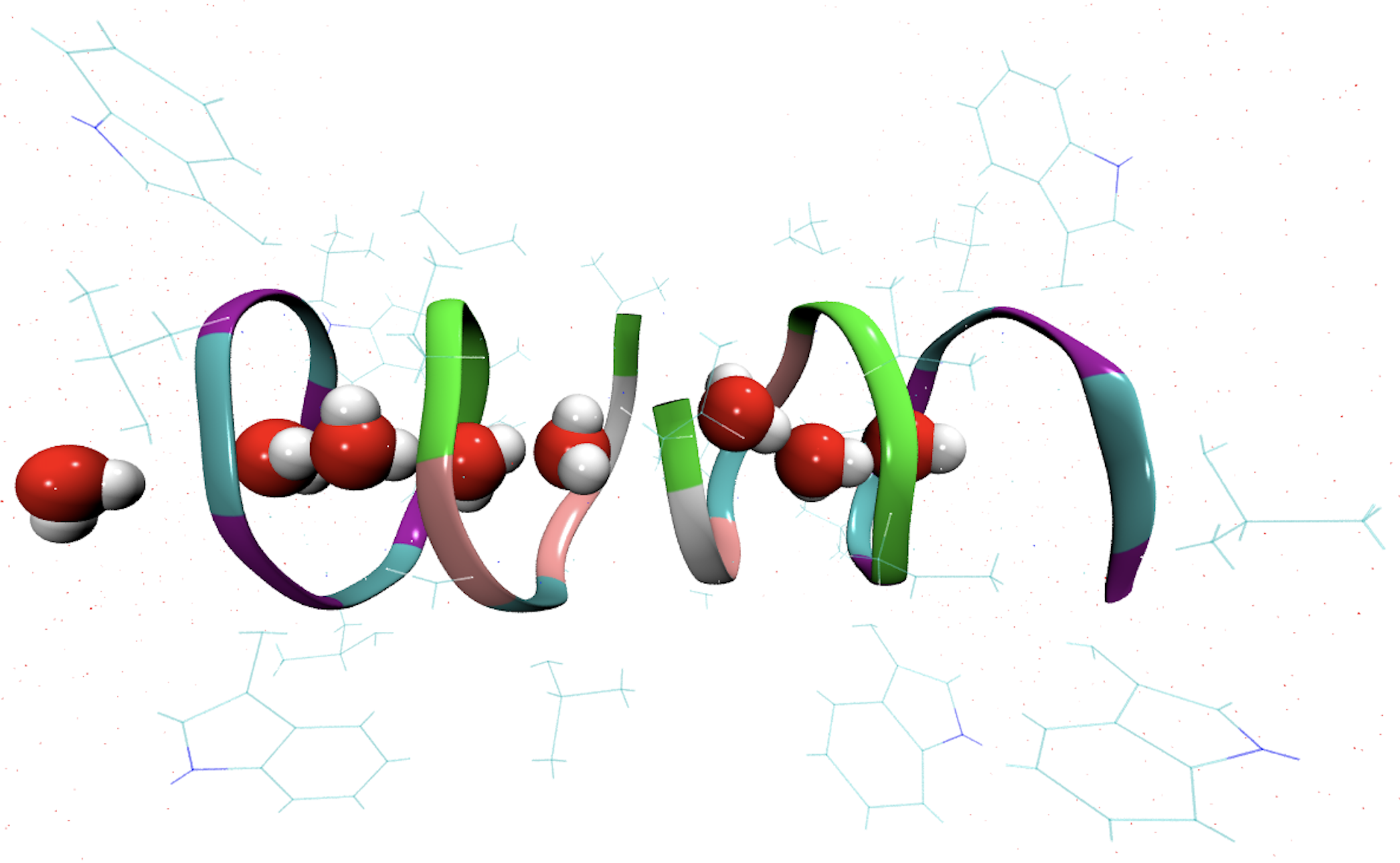}
    \caption{A water-wire confined within the Gramicidin ion channel.}
    \label{fig:gramicidin}
 \end{figure}

\subsection{``Grotthuss''-type coupled stretch in protonated pentamer water wire} 
\label{pentamer}
In biological systems, the water wire described above is often partially confined within an active site or inside an ion-channel as illustrated in Fig.~\ref{fig:gramicidin}. In this section, we analyze an instance of such a system consisting of four degrees of freedom.

 Deferring the confinement modeling for a moment, and focusing on the four degrees of freedom, the matrix product state for the couple wavefunction has the form, 
\begin{multline}
   \psi(x_1, x_2, x_3, x_4) = \\
\sum_{i_1,i_2,i_3} \psi_{i_1}^1(x_1)\alpha_{i_1}^1  \psi_{i_1,i_2}^{2}(x_2) \alpha_{i_2}^2 \psi_{i_2,i_3}^{3}(x_3) \alpha_{i_3}^3 \psi_{i_3}^{4}(x_4) 
    \label{4D_TN_eqn}
\end{multline}
The quantities $\alpha_{i_j}^j$ in the equation above represent weights for the bond dimensions.

As done in the previous section, we can model various interrogation scenarios. We show the result of performing a measurement on dimension $R_1$. This reduces Eq.~(\ref{4D_TN_eqn}) to produce
\begin{multline}
  \bra{\psi_{k}^1(x_1)} \ket{\psi(x_1, x_2, x_3, x_4)} = \\
 \sum_{i_2,i_3} \alpha_{k}^1  \psi_{k,i_2}^{2}(x_2) \alpha_{i_2}^2 \psi_{i_2,i_3}^{3}(x_3)  
      \alpha_{i_3}^3 \psi_{i_3}^{4}(x_4) 
    \label{4D_TN_eqn-proj}
\end{multline}
and the corresponding measurement outcome is simply the magnitude of the vector in Eq.~(\ref{4D_TN_eqn-proj}): 
\begin{align}
  \bra{\psi(x_1, x_2, x_3, x_4)} \ket{\psi_{k}^1(x_1)} \bra{\psi_{k}^1(x_1)} \ket{\psi(x_1, x_2, x_3, x_4)} 
    \label{4D_TN_eqn-proj-prob}
\end{align}

\begin{figure*}[t]
\begin{center}
\includegraphics[width=0.7\columnwidth]{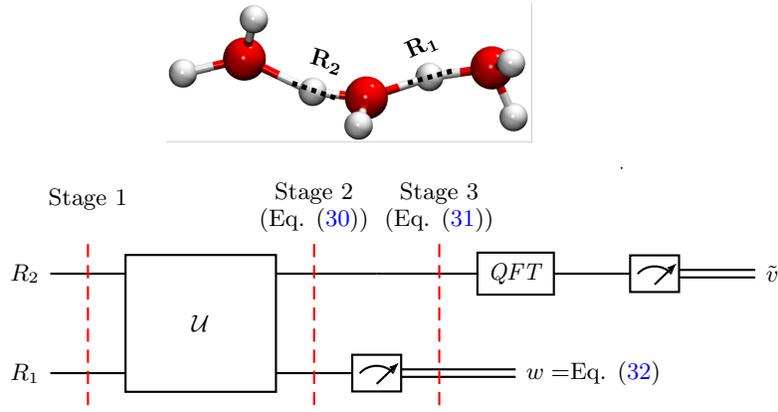} \\
\begin{quantikz}[row sep=0.7cm,column sep=1cm]
   \lstick{$R_2$}\slice{Stage 1\\} & 
   \gate[wires=2][2cm]{{\cal U}}\slice{Stage 2 \\ (Eq. (\ref{2D_TN_eqn}))} &
   \qw\slice{Stage 3 \\ (Eq. (\ref{2D_TN_eqn-proj}))} &
   \gate{\mathit{QFT}} &
   \meter{} & 
   \rstick{$\tilde{v}$}\cw
   \\
   \lstick{$R_1$} &
   &
   \meter{} &
   \rstick{$w = $Eq. (\ref{2D_TN_eqn-proj-prob})}\cw 
\end{quantikz} \\
\end{center}
\caption{\label{fig:shor-2Dwater}Quantum circuit version for protonated water-dimer system discussed in Section \ref{dimer}.}
\end{figure*}
\begin{figure*}[t]
\includegraphics[width=\columnwidth]{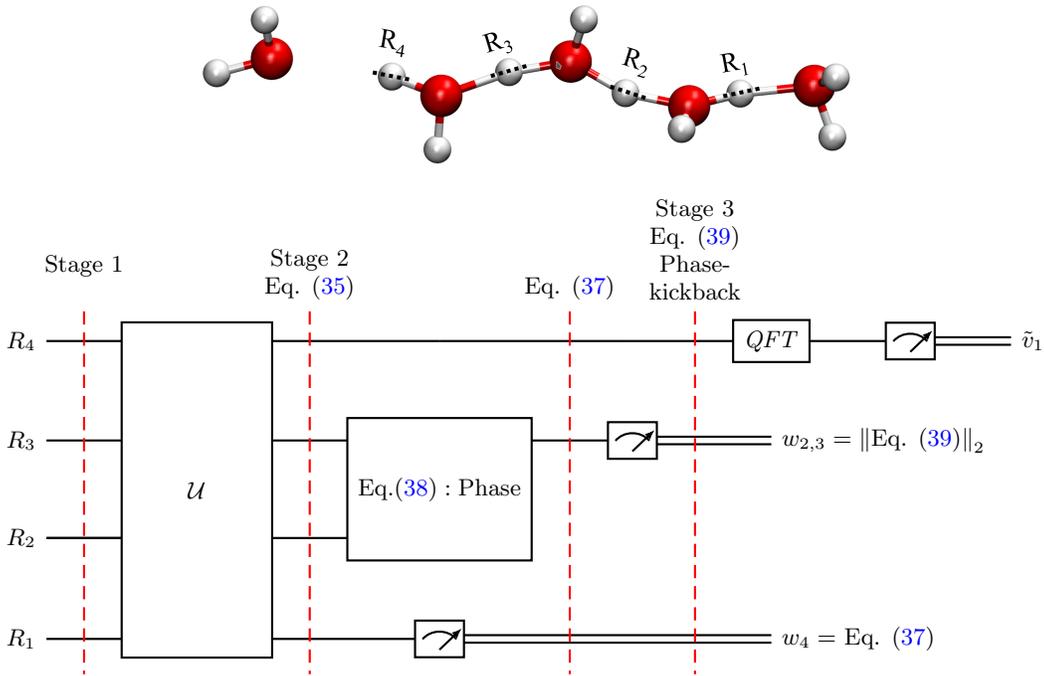}
\begin{center}
\begin{quantikz}[row sep=0.7cm,column sep=1cm]
   \lstick{$R_4$}\slice{Stage 1 \\ } & 
   \gate[wires=4][2cm]{ {\cal U} }\slice{Stage 2 \\ Eq. (\ref{4D_TN_eqn})} &
   \qw\slice{Eq. (\ref{4D_TN_eqn-proj-prob})} &  
   \qw\slice{Stage 3 \\ Eq. (\ref{4D_TN_eqn-proj-3}) \\ Phase-\\ kickback}   &
   \gate[wires=1]{\mathit{QFT}} &
   \meter{} & 
   \rstick{$\tilde{v}_1$}\cw
   \\
   \lstick{$R_3$} & & 
   \gate[wires=2]{{\text {Eq.} (\ref{4D_TN_eqn-proj-2}):  
   {\text {Phase}}}} & 
   \meter{} &
   \rstick{$w_{2,3} = {\left\| {\text {Eq. }} (\ref{4D_TN_eqn-proj-3}) \right\|}_2$}\cw
   \\
   \lstick{$R_2$} &
   \qw & & & \\
   \lstick{$R_1$} &
   &
   \meter{} & \cw &
   \rstick{$w_4 = $ Eq. (\ref{4D_TN_eqn-proj-prob})}\cw 
\end{quantikz}
\end{center}
\caption{\label{fig:shor-4Dwater}Quantum circuit version for protonated water-pentamer system discussed in Section \ref{pentamer}. A combination of a simple measurement at $R_1$, given by Eqs. (\ref{4D_TN_eqn-proj}), (\ref{4D_TN_eqn-proj-prob}) and a rotation for $R_2,R_3$, given by Eq. (\ref{4D_TN_eqn-proj-3}) and its norm, result in a composite final outcome at Stage 3 for $R_4$.}
\end{figure*}

This simple analysis ignored the fact that in most biological systems (especially ion-channels as well as enzyme active sites), the water molecules that encapsulate dimensions $R_2$ and $R_3$ have limited degree of flexibility. This results in a limited degree of projection of dimensions $R_2$ and $R_3$ into, for example, a subspace given by
\begin{align}
   \chi(x_2,x_3) \equiv \sum_{i_2,i_3} \beta_{i_2,i_3} \psi_{k,i_2}^{2}(x_2) \psi_{i_2,i_3}^{3}(x_3)
    \label{4D_TN_eqn-proj-2}
\end{align}
Note that such a state is not dissimilar to the measurement basis state used in the phase kickback scenario at the end of Sec.~\ref{QCD-Shor} and reduces the possible outcomes for $x_4$. Thus, after projection of Eq.~(\ref{4D_TN_eqn-proj-2}) onto Eq.~(\ref{4D_TN_eqn-proj}), we obtain
\begin{align}
  \sum_{i_2,i_3} & \left( \beta_{i_2,i_3} \alpha_{k}^1  \alpha_{i_2}^2 \alpha_{i_3}^3 \right) \psi_{i_3}^{4}(x_4) 
    \label{4D_TN_eqn-proj-3}
\end{align}
where the bracketed set acts as a combined coefficient that curtails the set of possibilities for $x_4$. Thus, based on the extent of flexibility provided by the restrictions to $\beta_{i_2,i_3}$, there are a range of possible outcomes at the far end depicted by $x_4$. 

This idea can be generalized for an arbitrary number of degrees of freedom, where, using Eq.~(\ref{4D_TN_eqn-proj-3}), we may write our final result as
\begin{align}
  \sum_{i_2,\cdots,i_{N-1}} & \left( \beta_{i_2,\cdots,i_{N-1}} \alpha_{k}^1  \alpha_{i_2}^2 \cdots \alpha_{i_{N-1}}^{N-1} \right) \psi_{i_N}^{N}(x_N) 
    \label{4D_TN_eqn-proj-gen}
\end{align}

\subsection{Circuit model representation for the protonated water wire problems}

We now analyze the results from the two sections above using the quantum circuit model. The quantum circuits thus derived are based on Fig.~\ref{fig:shor-quantum} and are presented in Figs.~\ref{fig:shor-2Dwater} and~\ref{fig:shor-4Dwater}. As discussed in Section \ref{QCD-Shor-Circuit}, all parts of both systems are correlated at the end of Stage (2) with wavefunctions given by Eqs. (\ref{2D_TN_eqn}) and (\ref{4D_TN_eqn}). In the case of the protonated water dimer, a measurement along dimension $R_1$ follows, resulting in the projected state given by Eq. (\ref{2D_TN_eqn-proj}) at Stage (3), with measurement probability given by Eq. (\ref{2D_TN_eqn-proj-prob}). 

The protonated pentamer problem is complicated due to the phase-kickback step resulting from basis rotations (Eq. (\ref{4D_TN_eqn-proj-3}) on the $\left(R_2, R_3 \right)$ degrees of freedom as shown in Fig.~\ref{fig:shor-4Dwater}). Specifically, the correlated state at Stage (2) given by Eq. (\ref{4D_TN_eqn}) undergoes measurements at $R_1$, with outcome, Eq. (\ref{4D_TN_eqn-proj-prob}), where the corresponding resultant state is then projected onto a rotated basis within dimension $R_2$ and $R_3$, given by Eq. (\ref{4D_TN_eqn-proj-2}) to arrive at Stage (3) at state given by Eq. (\ref{4D_TN_eqn-proj-3}). Thus Figs.~\ref{fig:shor-2Dwater} and~\ref{fig:shor-4Dwater}, through the discussion accompanying Fig.~\ref{fig:shor-quantum}, provide a detailed analogue to the Shor algorithm in Fig.~\ref{fig:shor} by way of abstract formalism presented in Sec.~\ref{QCD-Shor}.

%%%%%%%%%%%%%%%%

%%%%%%%%%%%%%%%%%%%%%%%%%%%%%%%%%%%%%%%%%%%%%%%%%%%%%%%%%%%%%%%%%
\section{Conclusion}
\label{concl}

In this article, we have developed an abstract formalism of tensor networks based quantum dynamics applicable broadly for all quantum systems, and we discuss how this general idea captures the key signatures within the well-known Shor's algorithm as well as other well known quantum algorithms. In essence, for a general bipartite system, unitary evolution encodes within the evolved state, a characteristic correlation or entanglement that bears the signature of the Hamiltonian that defines the  evolution operator. Thus, once we interrogate a specific part of a quantum system, the remaining parts of that system automatically get projected based on the extent of correlation that resides within the system, and this is borne out of measurement. 

This close resemblance of the general formalism of quantum propagation of multi-partite systems, and the symmetry of integers as captured by Shor's algorithm, begs the rather profound question of whether natural systems exist, or can be designed, that may perform the operations that we might interpret as prime-factoring. We do not dwell into this general question in this article, but we ask the opposite question of whether we can exploit this connection to analyze quantum chemical dynamics. Indeed we find that, when we apply this concept along with the generalized definition of phase kickback as obtained from the tensor network description of Shor's algorithm and quantum dynamics described here, the coupled dynamics of protons in a protonated water wire naturally lends itself to a projected transport interpretation. We find that projection of states at one end of a water wire, along with phase kickback-like constrained operations along the length of the water chain provide a general description for proton transport that is commensurate with our description of Shor's algorithm. Future publications are currently planned to exploit this analogy to probe electron-nuclear dynamics in the non-adiabatic setting. Furthermore, given that the surrounding vibrational degrees of freedom may  be projected through phase-kickback as noted here, to tailor the dynamics within one mode ($R_4$ for the protonated pentamer), perhaps chemical catalysis is another area where these broad ideas may find application. These aspects will be probed in future work.

%%%%%%%%%%%%%%%%%%%%%%%%%%%%%%%%%%%%%%%%%%%%%%%%%%%%%%%%%%%%%%%%%
\section{Acknowledgments}

This research was supported by the National Science Foundation grants OMA-1936353 to authors SSI and AS.

%\bibliography{References/Quantum-Computing,References/TN-paper,References/waterrefs,References/water-clusters,References/srini-water,References/amr,References/math,References/qwaimdrefs,References/SSI-molecular-fragmentation}

\begin{thebibliography}{51}
\expandafter\ifx\csname natexlab\endcsname\relax\def\natexlab#1{#1}\fi
\expandafter\ifx\csname bibnamefont\endcsname\relax
  \def\bibnamefont#1{#1}\fi
\expandafter\ifx\csname bibfnamefont\endcsname\relax
  \def\bibfnamefont#1{#1}\fi
\expandafter\ifx\csname citenamefont\endcsname\relax
  \def\citenamefont#1{#1}\fi
\expandafter\ifx\csname url\endcsname\relax
  \def\url#1{\texttt{#1}}\fi
\expandafter\ifx\csname urlprefix\endcsname\relax\def\urlprefix{URL }\fi
\providecommand{\bibinfo}[2]{#2}
\providecommand{\eprint}[2][]{\url{#2}}

\bibitem[{\citenamefont{Preskill}(2018)}]{Preskill2018-NISQ}
\bibinfo{author}{\bibfnamefont{J.}~\bibnamefont{Preskill}},
  \bibinfo{journal}{{Quantum}} \textbf{\bibinfo{volume}{2}},
  \bibinfo{pages}{79} (\bibinfo{year}{2018}).

\bibitem[{\citenamefont{Aspuru-Guzik et~al.}({2005})\citenamefont{Aspuru-Guzik,
  Dutoi, Love, and Head-Gordon}}]{Aspuru-Guzik-Science-2005}
\bibinfo{author}{\bibfnamefont{A.}~\bibnamefont{Aspuru-Guzik}},
  \bibinfo{author}{\bibfnamefont{A.~D.} \bibnamefont{Dutoi}},
  \bibinfo{author}{\bibfnamefont{P.~J.} \bibnamefont{Love}}, \bibnamefont{and}
  \bibinfo{author}{\bibfnamefont{M.}~\bibnamefont{Head-Gordon}},
  \bibinfo{journal}{{Science}} \textbf{\bibinfo{volume}{{309}}},
  \bibinfo{pages}{{1704}} (\bibinfo{year}{{2005}}).

\bibitem[{\citenamefont{Shor}({1997})}]{Shor}
\bibinfo{author}{\bibfnamefont{P.~W.} \bibnamefont{Shor}},
  \bibinfo{journal}{{SIAM J. Comput. }} \textbf{\bibinfo{volume}{{26}}},
  \bibinfo{pages}{{1484–1509}} (\bibinfo{year}{{1997}}).

\bibitem[{\citenamefont{Nielsen and Chuang}(2000)}]{Nielsen-Chuang}
\bibinfo{author}{\bibfnamefont{M.~A.} \bibnamefont{Nielsen}} \bibnamefont{and}
  \bibinfo{author}{\bibfnamefont{I.~L.} \bibnamefont{Chuang}},
  \emph{\bibinfo{title}{Quantum computation and quantum information}}
  (\bibinfo{publisher}{Cambridge University Press}, \bibinfo{address}{New York,
  NY, USA}, \bibinfo{year}{2000}), ISBN \bibinfo{isbn}{0-521-63503-9}.

\bibitem[{\citenamefont{Or\'{u}s}(2014)}]{tensor_network}
\bibinfo{author}{\bibfnamefont{R.}~\bibnamefont{Or\'{u}s}},
  \bibinfo{journal}{Ann. Physics} \textbf{\bibinfo{volume}{349}},
  \bibinfo{pages}{117 } (\bibinfo{year}{2014}).

\bibitem[{\citenamefont{Oseledets}(2011)}]{TensorTrain}
\bibinfo{author}{\bibfnamefont{I.}~\bibnamefont{Oseledets}},
  \bibinfo{journal}{SIAM Journal on Scientific Computing}
  \textbf{\bibinfo{volume}{33}}, \bibinfo{pages}{2295} (\bibinfo{year}{2011}).

\bibitem[{\citenamefont{Kolda and Bader}(2009)}]{Kolda2009-sb}
\bibinfo{author}{\bibfnamefont{T.~G.} \bibnamefont{Kolda}} \bibnamefont{and}
  \bibinfo{author}{\bibfnamefont{B.~W.} \bibnamefont{Bader}},
  \bibinfo{journal}{SIAM Rev.} \textbf{\bibinfo{volume}{51}},
  \bibinfo{pages}{455} (\bibinfo{year}{2009}), ISSN \bibinfo{issn}{0036-1445}.

\bibitem[{\citenamefont{Verstraete et~al.}(2008)\citenamefont{Verstraete, Murg,
  and Cirac}}]{Verstraete2008-ah}
\bibinfo{author}{\bibfnamefont{F.}~\bibnamefont{Verstraete}},
  \bibinfo{author}{\bibfnamefont{V.}~\bibnamefont{Murg}}, \bibnamefont{and}
  \bibinfo{author}{\bibfnamefont{J.~I.} \bibnamefont{Cirac}},
  \bibinfo{journal}{Adv. Phys.} \textbf{\bibinfo{volume}{57}},
  \bibinfo{pages}{143} (\bibinfo{year}{2008}), ISSN \bibinfo{issn}{0001-8732}.

\bibitem[{\citenamefont{Bridgeman and Chubb}(2017)}]{Bridgeman2017-mh}
\bibinfo{author}{\bibfnamefont{J.~C.} \bibnamefont{Bridgeman}}
  \bibnamefont{and} \bibinfo{author}{\bibfnamefont{C.~T.} \bibnamefont{Chubb}},
  \bibinfo{journal}{J. Phys. A: Math. Theor.} \textbf{\bibinfo{volume}{50}},
  \bibinfo{pages}{223001} (\bibinfo{year}{2017}), ISSN
  \bibinfo{issn}{1751-8121}.

\bibitem[{\citenamefont{Montangero}(2018)}]{Montangero2018-ct}
\bibinfo{author}{\bibfnamefont{S.}~\bibnamefont{Montangero}},
  \emph{\bibinfo{title}{Introduction to Tensor Network Methods: Numerical
  simulations of low-dimensional many-body quantum systems}}
  (\bibinfo{publisher}{Springer, Cham}, \bibinfo{year}{2018}).

\bibitem[{\citenamefont{Ran et~al.}(2020)\citenamefont{Ran, Tirrito, Peng,
  Chen, Tagliacozzo, Su, and Lewenstein}}]{Ran2020-ve}
\bibinfo{author}{\bibfnamefont{S.-J.} \bibnamefont{Ran}},
  \bibinfo{author}{\bibfnamefont{E.}~\bibnamefont{Tirrito}},
  \bibinfo{author}{\bibfnamefont{C.}~\bibnamefont{Peng}},
  \bibinfo{author}{\bibfnamefont{X.}~\bibnamefont{Chen}},
  \bibinfo{author}{\bibfnamefont{L.}~\bibnamefont{Tagliacozzo}},
  \bibinfo{author}{\bibfnamefont{G.}~\bibnamefont{Su}}, \bibnamefont{and}
  \bibinfo{author}{\bibfnamefont{M.}~\bibnamefont{Lewenstein}},
  \emph{\bibinfo{title}{Tensor Network Contractions: Methods and Applications
  to Quantum {Many-Body} Systems}} (\bibinfo{publisher}{Springer, Cham},
  \bibinfo{year}{2020}).

\bibitem[{\citenamefont{Silvi et~al.}(2019)\citenamefont{Silvi, Tschirsich,
  Gerster, J{\"u}nemann, Jaschke, Rizzi, and Montangero}}]{Silvi2019-hl}
\bibinfo{author}{\bibfnamefont{P.}~\bibnamefont{Silvi}},
  \bibinfo{author}{\bibfnamefont{F.}~\bibnamefont{Tschirsich}},
  \bibinfo{author}{\bibfnamefont{M.}~\bibnamefont{Gerster}},
  \bibinfo{author}{\bibfnamefont{J.}~\bibnamefont{J{\"u}nemann}},
  \bibinfo{author}{\bibfnamefont{D.}~\bibnamefont{Jaschke}},
  \bibinfo{author}{\bibfnamefont{M.}~\bibnamefont{Rizzi}}, \bibnamefont{and}
  \bibinfo{author}{\bibfnamefont{S.}~\bibnamefont{Montangero}},
  \bibinfo{journal}{SciPost Phys. Lect. Notes}  (\bibinfo{year}{2019}), ISSN
  \bibinfo{issn}{2590-1990}.

\bibitem[{\citenamefont{Or{\'u}s}(2019)}]{Orus2019-wn}
\bibinfo{author}{\bibfnamefont{R.}~\bibnamefont{Or{\'u}s}},
  \bibinfo{journal}{Nature Reviews Physics} \textbf{\bibinfo{volume}{1}},
  \bibinfo{pages}{538} (\bibinfo{year}{2019}), ISSN \bibinfo{issn}{2522-5820}.

\bibitem[{\citenamefont{Or{\'u}s}(2014)}]{Orus2014-gg}
\bibinfo{author}{\bibfnamefont{R.}~\bibnamefont{Or{\'u}s}},
  \bibinfo{journal}{Ann. Phys.} \textbf{\bibinfo{volume}{349}},
  \bibinfo{pages}{117} (\bibinfo{year}{2014}), ISSN \bibinfo{issn}{0003-4916}.

\bibitem[{\citenamefont{Biamonte and Bergholm}(2017)}]{Biamonte2017-xj}
\bibinfo{author}{\bibfnamefont{J.}~\bibnamefont{Biamonte}} \bibnamefont{and}
  \bibinfo{author}{\bibfnamefont{V.}~\bibnamefont{Bergholm}}
  (\bibinfo{year}{2017}), \eprint{1708.00006}.

\bibitem[{\citenamefont{Biamonte}(2019)}]{Biamonte2019-du}
\bibinfo{author}{\bibfnamefont{J.}~\bibnamefont{Biamonte}}
  (\bibinfo{year}{2019}), \eprint{1912.10049}.

\bibitem[{\citenamefont{Eisert}(2013)}]{Eisert2013-yn}
\bibinfo{author}{\bibfnamefont{J.}~\bibnamefont{Eisert}}
  (\bibinfo{year}{2013}), \eprint{1308.3318}.

\bibitem[{\citenamefont{Peng et~al.}(2020)\citenamefont{Peng, Harrow, Ozols,
  and Wu}}]{Qubit-fragmentation}
\bibinfo{author}{\bibfnamefont{T.}~\bibnamefont{Peng}},
  \bibinfo{author}{\bibfnamefont{A.~W.} \bibnamefont{Harrow}},
  \bibinfo{author}{\bibfnamefont{M.}~\bibnamefont{Ozols}}, \bibnamefont{and}
  \bibinfo{author}{\bibfnamefont{X.}~\bibnamefont{Wu}}, \bibinfo{journal}{Phys.
  Rev. Lett.} \textbf{\bibinfo{volume}{125}}, \bibinfo{pages}{150504}
  (\bibinfo{year}{2020}),
  \urlprefix\url{https://link.aps.org/doi/10.1103/PhysRevLett.125.150504}.

\bibitem[{\citenamefont{Fried et~al.}(2018)\citenamefont{Fried, Sawaya, Cao,
  Kivlichan, Romero, and Aspuru-Guzik}}]{Fried2018-th}
\bibinfo{author}{\bibfnamefont{E.~S.} \bibnamefont{Fried}},
  \bibinfo{author}{\bibfnamefont{N.~P.~D.} \bibnamefont{Sawaya}},
  \bibinfo{author}{\bibfnamefont{Y.}~\bibnamefont{Cao}},
  \bibinfo{author}{\bibfnamefont{I.~D.} \bibnamefont{Kivlichan}},
  \bibinfo{author}{\bibfnamefont{J.}~\bibnamefont{Romero}}, \bibnamefont{and}
  \bibinfo{author}{\bibfnamefont{A.}~\bibnamefont{Aspuru-Guzik}},
  \bibinfo{journal}{PLoS One} \textbf{\bibinfo{volume}{13}},
  \bibinfo{pages}{e0208510} (\bibinfo{year}{2018}), ISSN
  \bibinfo{issn}{1932-6203},
  \urlprefix\url{http://dx.doi.org/10.1371/journal.pone.0208510}.

\bibitem[{\citenamefont{Schutski et~al.}(2020)\citenamefont{Schutski, Lykov,
  and Oseledets}}]{Schutski2020-rp}
\bibinfo{author}{\bibfnamefont{R.}~\bibnamefont{Schutski}},
  \bibinfo{author}{\bibfnamefont{D.}~\bibnamefont{Lykov}}, \bibnamefont{and}
  \bibinfo{author}{\bibfnamefont{I.}~\bibnamefont{Oseledets}},
  \bibinfo{journal}{Phys. Rev. A} \textbf{\bibinfo{volume}{101}},
  \bibinfo{pages}{042335} (\bibinfo{year}{2020}), ISSN
  \bibinfo{issn}{1050-2947},
  \urlprefix\url{https://link.aps.org/doi/10.1103/PhysRevA.101.042335}.

\bibitem[{\citenamefont{Bauer et~al.}(2020)\citenamefont{Bauer, Bravyi, Motta,
  and Kin-Lic~Chan}}]{Bauer2020-cb}
\bibinfo{author}{\bibfnamefont{B.}~\bibnamefont{Bauer}},
  \bibinfo{author}{\bibfnamefont{S.}~\bibnamefont{Bravyi}},
  \bibinfo{author}{\bibfnamefont{M.}~\bibnamefont{Motta}}, \bibnamefont{and}
  \bibinfo{author}{\bibfnamefont{G.}~\bibnamefont{Kin-Lic~Chan}},
  \bibinfo{journal}{Chem. Rev.} \textbf{\bibinfo{volume}{120}},
  \bibinfo{pages}{12685} (\bibinfo{year}{2020}), ISSN \bibinfo{issn}{0009-2665,
  1520-6890}, \urlprefix\url{http://dx.doi.org/10.1021/acs.chemrev.9b00829}.

\bibitem[{\citenamefont{White}(1992)}]{DMRG-White}
\bibinfo{author}{\bibfnamefont{S.~R.} \bibnamefont{White}},
  \bibinfo{journal}{Phys. Rev. Lett.} \textbf{\bibinfo{volume}{69}},
  \bibinfo{pages}{2863} (\bibinfo{year}{1992}).

\bibitem[{\citenamefont{Schollw\"ock}(2011)}]{DMRG_MPS}
\bibinfo{author}{\bibfnamefont{U.}~\bibnamefont{Schollw\"ock}},
  \bibinfo{journal}{Ann. Physics} \textbf{\bibinfo{volume}{326}},
  \bibinfo{pages}{96 } (\bibinfo{year}{2011}), ISSN \bibinfo{issn}{0003-4916},
  \bibinfo{note}{january 2011 Special Issue}.

\bibitem[{\citenamefont{Schollw\"{o}ck}(2005)}]{DMRG}
\bibinfo{author}{\bibfnamefont{U.}~\bibnamefont{Schollw\"{o}ck}},
  \bibinfo{journal}{Rev. Mod. Phys.} \textbf{\bibinfo{volume}{77}},
  \bibinfo{pages}{259} (\bibinfo{year}{2005}).

\bibitem[{\citenamefont{Keller and Reiher}(2014)}]{DMRGCI1}
\bibinfo{author}{\bibfnamefont{S.~F.} \bibnamefont{Keller}} \bibnamefont{and}
  \bibinfo{author}{\bibfnamefont{M.}~\bibnamefont{Reiher}},
  \bibinfo{journal}{CHIMA International Journal for Chemistry}
  \textbf{\bibinfo{volume}{68}}, \bibinfo{pages}{200} (\bibinfo{year}{2014}).

\bibitem[{\citenamefont{Vidal}(2004)}]{Vidal2004-oh}
\bibinfo{author}{\bibfnamefont{G.}~\bibnamefont{Vidal}},
  \bibinfo{journal}{Phys. Rev. Lett.} \textbf{\bibinfo{volume}{93}},
  \bibinfo{pages}{040502} (\bibinfo{year}{2004}), ISSN
  \bibinfo{issn}{0031-9007}.

\bibitem[{\citenamefont{Szalay et~al.}(2015)\citenamefont{Szalay, Pfeffer,
  Murg, Barcza, Verstraete, Schneider, and Legeza}}]{TNabinitio}
\bibinfo{author}{\bibfnamefont{S.}~\bibnamefont{Szalay}},
  \bibinfo{author}{\bibfnamefont{M.}~\bibnamefont{Pfeffer}},
  \bibinfo{author}{\bibfnamefont{V.}~\bibnamefont{Murg}},
  \bibinfo{author}{\bibfnamefont{G.}~\bibnamefont{Barcza}},
  \bibinfo{author}{\bibfnamefont{F.}~\bibnamefont{Verstraete}},
  \bibinfo{author}{\bibfnamefont{R.}~\bibnamefont{Schneider}},
  \bibnamefont{and} \bibinfo{author}{\bibfnamefont{O.}~\bibnamefont{Legeza}},
  \bibinfo{journal}{Int. J. Quant. Chem.} \textbf{\bibinfo{volume}{115}},
  \bibinfo{pages}{1342} (\bibinfo{year}{2015}).

\bibitem[{\citenamefont{Chan et~al.}(2016)\citenamefont{Chan, Keselman,
  Nakatani, Li, and White}}]{Chan-While-MPS-MPO}
\bibinfo{author}{\bibfnamefont{G.~K.-L.} \bibnamefont{Chan}},
  \bibinfo{author}{\bibfnamefont{A.}~\bibnamefont{Keselman}},
  \bibinfo{author}{\bibfnamefont{N.}~\bibnamefont{Nakatani}},
  \bibinfo{author}{\bibfnamefont{Z.}~\bibnamefont{Li}}, \bibnamefont{and}
  \bibinfo{author}{\bibfnamefont{S.~R.} \bibnamefont{White}},
  \bibinfo{journal}{J. Chem. Phys.} \textbf{\bibinfo{volume}{145}},
  \bibinfo{pages}{014102} (\bibinfo{year}{2016}).

\bibitem[{\citenamefont{Gunst et~al.}(2018)\citenamefont{Gunst, Verstraete,
  Wouters, Legeza, and Van~Neck}}]{T3NS}
\bibinfo{author}{\bibfnamefont{K.}~\bibnamefont{Gunst}},
  \bibinfo{author}{\bibfnamefont{F.}~\bibnamefont{Verstraete}},
  \bibinfo{author}{\bibfnamefont{S.}~\bibnamefont{Wouters}},
  \bibinfo{author}{\bibfnamefont{O.}~\bibnamefont{Legeza}}, \bibnamefont{and}
  \bibinfo{author}{\bibfnamefont{D.}~\bibnamefont{Van~Neck}},
  \bibinfo{journal}{J. Chem. Theory and Comput.} \textbf{\bibinfo{volume}{14}},
  \bibinfo{pages}{2026} (\bibinfo{year}{2018}).

\bibitem[{\citenamefont{Kumar et~al.}(2022)\citenamefont{Kumar, DeGregorio,
  Ricard, and Iyengar}}]{frag-TN-Anup}
\bibinfo{author}{\bibfnamefont{A.}~\bibnamefont{Kumar}},
  \bibinfo{author}{\bibfnamefont{N.}~\bibnamefont{DeGregorio}},
  \bibinfo{author}{\bibfnamefont{T.}~\bibnamefont{Ricard}}, \bibnamefont{and}
  \bibinfo{author}{\bibfnamefont{S.~S.} \bibnamefont{Iyengar}},
  \bibinfo{journal}{J. Chem. Theory Comput.} \textbf{\bibinfo{volume}{18}},
  \bibinfo{pages}{7243} (\bibinfo{year}{2022}).

\bibitem[{\citenamefont{Baranov and Oseledets}(2015)}]{AS+TT}
\bibinfo{author}{\bibfnamefont{V.}~\bibnamefont{Baranov}} \bibnamefont{and}
  \bibinfo{author}{\bibfnamefont{I.}~\bibnamefont{Oseledets}},
  \bibinfo{journal}{J. Chem. Phys.} \textbf{\bibinfo{volume}{143}},
  \bibinfo{pages}{174107} (\bibinfo{year}{2015}).

\bibitem[{\citenamefont{DeGregorio and Iyengar}(2019)}]{Nicole-TN}
\bibinfo{author}{\bibfnamefont{N.}~\bibnamefont{DeGregorio}} \bibnamefont{and}
  \bibinfo{author}{\bibfnamefont{S.~S.} \bibnamefont{Iyengar}},
  \bibinfo{journal}{J. Chem. Theory Comput.} \textbf{\bibinfo{volume}{15}},
  \bibinfo{pages}{2780} (\bibinfo{year}{2019}).

\bibitem[{\citenamefont{Rajwade et~al.}(2013)\citenamefont{Rajwade, Rangarajan,
  and Banerjee}}]{rajwade2013image}
\bibinfo{author}{\bibfnamefont{A.}~\bibnamefont{Rajwade}},
  \bibinfo{author}{\bibfnamefont{A.}~\bibnamefont{Rangarajan}},
  \bibnamefont{and} \bibinfo{author}{\bibfnamefont{A.}~\bibnamefont{Banerjee}},
  \bibinfo{journal}{IEEE Transactions on Pattern Analysis and Machine
  Intelligence} \textbf{\bibinfo{volume}{35}}, \bibinfo{pages}{849}
  (\bibinfo{year}{2013}).

\bibitem[{\citenamefont{Hu et~al.}(2017)\citenamefont{Hu, Lu, Ye, and
  Zeng}}]{SVD_Noise_3}
\bibinfo{author}{\bibfnamefont{C.}~\bibnamefont{Hu}},
  \bibinfo{author}{\bibfnamefont{X.}~\bibnamefont{Lu}},
  \bibinfo{author}{\bibfnamefont{M.}~\bibnamefont{Ye}}, \bibnamefont{and}
  \bibinfo{author}{\bibfnamefont{W.}~\bibnamefont{Zeng}},
  \bibinfo{journal}{Pattern Recognition} \textbf{\bibinfo{volume}{64}},
  \bibinfo{pages}{60 } (\bibinfo{year}{2017}), ISSN \bibinfo{issn}{0031-3203}.

\bibitem[{\citenamefont{Iqbal et~al.}(2016)\citenamefont{Iqbal, Zerguine, Kaka,
  and Al-Shuhail}}]{SVD_Noise_2}
\bibinfo{author}{\bibfnamefont{N.}~\bibnamefont{Iqbal}},
  \bibinfo{author}{\bibfnamefont{A.}~\bibnamefont{Zerguine}},
  \bibinfo{author}{\bibfnamefont{S.}~\bibnamefont{Kaka}}, \bibnamefont{and}
  \bibinfo{author}{\bibfnamefont{A.}~\bibnamefont{Al-Shuhail}},
  \bibinfo{journal}{Journal of Geophysics and Engineering}
  \textbf{\bibinfo{volume}{13}}, \bibinfo{pages}{964} (\bibinfo{year}{2016}).

\bibitem[{\citenamefont{Hassanpour}(2008)}]{SVD_Noise_1}
\bibinfo{author}{\bibfnamefont{H.}~\bibnamefont{Hassanpour}},
  \bibinfo{journal}{Digital Signal Processing} \textbf{\bibinfo{volume}{18}},
  \bibinfo{pages}{728 } (\bibinfo{year}{2008}), ISSN \bibinfo{issn}{1051-2004}.

\bibitem[{\citenamefont{Kanjilal et~al.}(1997)\citenamefont{Kanjilal, Palit,
  and Saha}}]{SVR}
\bibinfo{author}{\bibfnamefont{P.~P.} \bibnamefont{Kanjilal}},
  \bibinfo{author}{\bibfnamefont{S.}~\bibnamefont{Palit}}, \bibnamefont{and}
  \bibinfo{author}{\bibfnamefont{G.}~\bibnamefont{Saha}},
  \bibinfo{journal}{IEEE Transactions on Biomedical Engineering}
  \textbf{\bibinfo{volume}{44}}, \bibinfo{pages}{51} (\bibinfo{year}{1997}).

\bibitem[{\citenamefont{Bernstein and
  Vazirani}(1997)}]{doi:10.1137/S0097539796300921}
\bibinfo{author}{\bibfnamefont{E.}~\bibnamefont{Bernstein}} \bibnamefont{and}
  \bibinfo{author}{\bibfnamefont{U.}~\bibnamefont{Vazirani}},
  \bibinfo{journal}{SIAM Journal on Computing} \textbf{\bibinfo{volume}{26}},
  \bibinfo{pages}{1411} (\bibinfo{year}{1997}),
  \eprint{https://doi.org/10.1137/S0097539796300921},
  \urlprefix\url{https://doi.org/10.1137/S0097539796300921}.

\bibitem[{\citenamefont{Deutsch}(1985)}]{deutsch}
\bibinfo{author}{\bibfnamefont{D.}~\bibnamefont{Deutsch}},
  \bibinfo{journal}{Proc. R. Soc. Lond. A 400}  (\bibinfo{year}{1985}).

\bibitem[{\citenamefont{Deutsch and Jozsa}(1992)}]{deutschJozsa}
\bibinfo{author}{\bibfnamefont{D.}~\bibnamefont{Deutsch}} \bibnamefont{and}
  \bibinfo{author}{\bibfnamefont{R.}~\bibnamefont{Jozsa}},
  \bibinfo{journal}{Proc. R. Soc. Lond. A 439}  (\bibinfo{year}{1992}).

\bibitem[{\citenamefont{Simon}(1994)}]{365701}
\bibinfo{author}{\bibfnamefont{D.}~\bibnamefont{Simon}}, in
  \emph{\bibinfo{booktitle}{Proceedings 35th Annual Symposium on Foundations of
  Computer Science}} (\bibinfo{year}{1994}), pp. \bibinfo{pages}{116--123}.

\bibitem[{\citenamefont{Shor}(1997)}]{doi:10.1137/S0097539795293172}
\bibinfo{author}{\bibfnamefont{P.~W.} \bibnamefont{Shor}},
  \bibinfo{journal}{SIAM Journal on Computing} \textbf{\bibinfo{volume}{26}},
  \bibinfo{pages}{1484} (\bibinfo{year}{1997}),
  \eprint{https://doi.org/10.1137/S0097539795293172},
  \urlprefix\url{https://doi.org/10.1137/S0097539795293172}.

\bibitem[{\citenamefont{Nielsen and Chuang}(2010)}]{nielsen_chuang_2010}
\bibinfo{author}{\bibfnamefont{M.~A.} \bibnamefont{Nielsen}} \bibnamefont{and}
  \bibinfo{author}{\bibfnamefont{I.~L.} \bibnamefont{Chuang}},
  \emph{\bibinfo{title}{Quantum Computation and Quantum Information: 10th
  Anniversary Edition}} (\bibinfo{publisher}{Cambridge University Press},
  \bibinfo{year}{2010}).

\bibitem[{\citenamefont{Grover}(1996)}]{10.1145/237814.237866}
\bibinfo{author}{\bibfnamefont{L.~K.} \bibnamefont{Grover}}, in
  \emph{\bibinfo{booktitle}{Proceedings of the Twenty-Eighth Annual ACM
  Symposium on Theory of Computing}} (\bibinfo{publisher}{Association for
  Computing Machinery}, \bibinfo{address}{New York, NY, USA},
  \bibinfo{year}{1996}), STOC '96, p. \bibinfo{pages}{212–219}, ISBN
  \bibinfo{isbn}{0897917855},
  \urlprefix\url{https://doi.org/10.1145/237814.237866}.

\bibitem[{\citenamefont{Coppersmith}(1994)}]{aqft}
\bibinfo{author}{\bibfnamefont{D.}~\bibnamefont{Coppersmith}},
  \bibinfo{journal}{IBM Research Report}  (\bibinfo{year}{1994}).

\bibitem[{\citenamefont{Dumitrescu}(2017)}]{Dumitrescu-TTN-QC}
\bibinfo{author}{\bibfnamefont{E.}~\bibnamefont{Dumitrescu}},
  \bibinfo{journal}{Phys. Rev. A} \textbf{\bibinfo{volume}{96}},
  \bibinfo{pages}{062322} (\bibinfo{year}{2017}),
  \urlprefix\url{https://link.aps.org/doi/10.1103/PhysRevA.96.062322}.

\bibitem[{\citenamefont{Vedral et~al.}(1996)\citenamefont{Vedral, Barenco, and
  Ekert}}]{PhysRevA.54.147}
\bibinfo{author}{\bibfnamefont{V.}~\bibnamefont{Vedral}},
  \bibinfo{author}{\bibfnamefont{A.}~\bibnamefont{Barenco}}, \bibnamefont{and}
  \bibinfo{author}{\bibfnamefont{A.}~\bibnamefont{Ekert}},
  \bibinfo{journal}{Phys. Rev. A} \textbf{\bibinfo{volume}{54}},
  \bibinfo{pages}{147} (\bibinfo{year}{1996}),
  \urlprefix\url{https://link.aps.org/doi/10.1103/PhysRevA.54.147}.

\bibitem[{\citenamefont{Schumaker et~al.}(2000)\citenamefont{Schumaker, Pomes,
  and Roux}}]{Schumaker:2000}
\bibinfo{author}{\bibfnamefont{M.~F.} \bibnamefont{Schumaker}},
  \bibinfo{author}{\bibfnamefont{R.}~\bibnamefont{Pomes}}, \bibnamefont{and}
  \bibinfo{author}{\bibfnamefont{B.}~\bibnamefont{Roux}},
  \bibinfo{journal}{Biophys.J.} \textbf{\bibinfo{volume}{79}},
  \bibinfo{pages}{2840} (\bibinfo{year}{2000}).

\bibitem[{\citenamefont{Schumaker et~al.}(2001)\citenamefont{Schumaker, Pomes,
  and Roux}}]{Schumaker:2001}
\bibinfo{author}{\bibfnamefont{M.~F.} \bibnamefont{Schumaker}},
  \bibinfo{author}{\bibfnamefont{R.}~\bibnamefont{Pomes}}, \bibnamefont{and}
  \bibinfo{author}{\bibfnamefont{B.}~\bibnamefont{Roux}},
  \bibinfo{journal}{Biophys.J.} \textbf{\bibinfo{volume}{80}},
  \bibinfo{pages}{12} (\bibinfo{year}{2001}).

\bibitem[{\citenamefont{Shin et~al.}(2004)\citenamefont{Shin, Hammer, Diken,
  Johnson, Walters, Jaeger, Duncan, Christie, and
  Jordan}}]{Johnson-Jordan-21mer}
\bibinfo{author}{\bibfnamefont{J.-W.} \bibnamefont{Shin}},
  \bibinfo{author}{\bibfnamefont{N.~I.} \bibnamefont{Hammer}},
  \bibinfo{author}{\bibfnamefont{E.~G.} \bibnamefont{Diken}},
  \bibinfo{author}{\bibfnamefont{M.~A.} \bibnamefont{Johnson}},
  \bibinfo{author}{\bibfnamefont{R.~S.} \bibnamefont{Walters}},
  \bibinfo{author}{\bibfnamefont{T.~D.} \bibnamefont{Jaeger}},
  \bibinfo{author}{\bibfnamefont{M.~A.} \bibnamefont{Duncan}},
  \bibinfo{author}{\bibfnamefont{R.~A.} \bibnamefont{Christie}},
  \bibnamefont{and} \bibinfo{author}{\bibfnamefont{K.~D.}
  \bibnamefont{Jordan}}, \bibinfo{journal}{Science}
  \textbf{\bibinfo{volume}{304}}, \bibinfo{pages}{1137} (\bibinfo{year}{2004}).

\bibitem[{\citenamefont{Iyengar et~al.}(2005)\citenamefont{Iyengar, Petersen,
  Day, Burnham, Teige, and Voth}}]{admp-21mer}
\bibinfo{author}{\bibfnamefont{S.~S.} \bibnamefont{Iyengar}},
  \bibinfo{author}{\bibfnamefont{M.~K.} \bibnamefont{Petersen}},
  \bibinfo{author}{\bibfnamefont{T.~J.~F.} \bibnamefont{Day}},
  \bibinfo{author}{\bibfnamefont{C.~J.} \bibnamefont{Burnham}},
  \bibinfo{author}{\bibfnamefont{V.~E.} \bibnamefont{Teige}}, \bibnamefont{and}
  \bibinfo{author}{\bibfnamefont{G.~A.} \bibnamefont{Voth}},
  \bibinfo{journal}{J. Chem. Phys.} \textbf{\bibinfo{volume}{123}},
  \bibinfo{pages}{084309} (\bibinfo{year}{2005}).

\end{thebibliography}

\end{document}